\documentclass[iop,numberedappendix,appendixfloats,twocolappendix,revtex4]{emulateapj}

\usepackage{amsmath, latexsym, color, verbatim}
\usepackage{graphicx,epsfig, natbib, epstopdf}
\usepackage{pdflscape, rotating}

\def\alwaysmath#1{\ifmmode{#1}\else{$#1$}\fi}
\shorttitle{SDSS-IV MaNGA: Ionized Gas Velocity Dispersions}
\shortauthors{Law et al. 2022}

\newcommand{\Ha}{\ensuremath{\rm H\alpha}}
\newcommand{\Hb}{\ensuremath{\rm H\beta}}

\newcommand{\kms}{\textrm{km~s}\ensuremath{^{-1}\,}}

\newcommand{\oone}{\textrm{[O\,{\sc i}]}}

\newcommand{\othree}{\textrm{[O\,{\sc iii}]}}
\newcommand{\otwo}{\textrm{[O\,{\sc ii}]}}
\newcommand{\ntwo}{\textrm{[N\,{\sc ii}]}}
\newcommand{\stwo}{\textrm{[S\,{\sc ii}]}}
\newcommand{\ctwo}{\textrm{[C\,{\sc ii}]}}

\begin{document}


\title{SDSS-IV MaNGA: Understanding Ionized Gas Turbulence using Integral Field Spectroscopy of 4500 Star-Forming Disk Galaxies}

\author{David R.~Law\altaffilmark{1},
Francesco Belfiore\altaffilmark{2},
Matthew A.~Bershady\altaffilmark{3,4,5}, 
Michele Cappellari\altaffilmark{6},
Niv Drory\altaffilmark{7},
Karen L.~Masters\altaffilmark{8},
Kyle B.~Westfall\altaffilmark{9},
Dmitry Bizyaev\altaffilmark{10},
Kevin Bundy\altaffilmark{9},
Kaike Pan\altaffilmark{10},
Renbin Yan\altaffilmark{11}
}

\altaffiltext{1}{Space Telescope Science Institute, 3700 San Martin Drive, Baltimore, MD 21218, USA; dlaw@stsci.edu}
\altaffiltext{2}{INAF -- Osservatorio Astrofisico di Arcetri, Largo E. Fermi 5, I-50157, Firenze, Italy.}
\altaffiltext{3}{University of Wisconsin - Madison, Department of Astronomy, 475 N. Charter Street, Madison, WI 53706-1582, USA.}
\altaffiltext{4}{South African Astronomical Observatory, PO Box 9, Observatory 7935, Cape Town, South Africa.}
\altaffiltext{5}{Department of Astronomy, University of Cape Town, Private Bag X3, Rondebosch 7701, South Africa.}
\altaffiltext{6}{Sub-department of Astrophysics, Department of Physics, University of Oxford, Denys Wilkinson Building, Keble Road, Oxford OX1 3RH, UK.}
\altaffiltext{7}{McDonald Observatory, The University of Texas at Austin, 2515 Speedway, Stop C1402, Austin, TX 78712, USA.}
\altaffiltext{8}{Departments of Physics and Astronomy, Haverford College, 370 Lancaster Avenue, Haverford, Pennsylvania 19041, USA.}
\altaffiltext{9}{University of California Observatories - Lick Observatory, University of California Santa Cruz, 1156 High St., Santa Cruz, CA 95064, USA.}
\altaffiltext{10}{Apache Point Observatory and New Mexico State
University, P.O. Box 59, Sunspot, NM, 88349-0059, USA}
\altaffiltext{11}{Department of Physics, The Chinese University of Hong Kong, Shatin, N.T., Hong Kong S.A.R., China.}

\begin{abstract}

The Sloan Digital Sky Survey MaNGA program has now obtained integral field spectroscopy for over 10,000 galaxies
in the nearby universe.  
We use the final MaNGA data release DR17 to study the correlation between ionized gas velocity dispersion and galactic star formation
rate, finding a tight correlation in which
$\sigma_{\Ha}$ from galactic HII regions increases significantly from $\sim 18-30$ \kms\
broadly in keeping with previous studies.
In contrast, $\sigma_{\Ha}$ from diffuse ionized gas (DIG) increases more rapidly from $20-60$ \kms.
Using the statistical power of MaNGA, we investigate these correlations in greater detail using multiple emission lines
and determine that the observed correlation of $\sigma_{\Ha}$ with local star formation rate surface density is
driven primarily by the global relation of increasing velocity dispersion at higher total SFR, as are apparent correlations with stellar mass.
Assuming HII region models consistent with our finding that $\sigma_{\othree} < \sigma_{\Ha} < \sigma_{\oone}$, we estimate the velocity dispersion of
the molecular gas in which individual HII regions are embedded, finding values $\sigma_{\rm Mol} = 5-30$ \kms\ consistent with ALMA observations
in a similar mass range.
Finally, we use variations in the relation with inclination and disk azimuthal angle to
constrain the velocity dispersion ellipsoid of the ionized gas $\sigma_z/\sigma_r = 0.84 \pm 0.03$ and
$\sigma_{\phi}/\sigma_r = 0.91 \pm 0.03$, similar to that of young stars in the Galactic disk.
Our results are most consistent with theoretical models in which turbulence in
modern galactic disks is driven primarily by star formation feedback.

\end{abstract}



\section{Introduction}

The gaseous interstellar medium (ISM) extends throughout galaxies, the densest molecular phase of which is closely
confined to the midplane of the galactic disk.  The distribution of this gas is far from uniform though; periodically localized overdensities
become sufficiently massive that their gravity can overcome hydrostatic gas pressure and trigger bursts of star formation.  The bright O and B stars
produced by this star formation 
ionize the surrounding ISM on the timescale of a few Myr, illuminating these regions in nebular emission lines such as
\Ha, \ntwo $\lambda6585$, and \othree\ $\lambda5007$ that can be observed to cosmological distances.  On small scales, the gas velocity dispersion observed in such emission lines can tell us about the physical conditions
within individual HII regions.  On large scales $\sim$ a few hundred pc to a kpc, such dispersions also encode information about the overall structure of the molecular gas
disk in which the HII regions are embedded; the corresponding turbulence is thus tightly linked to the overall stability of the gaseous disk.

One of the major puzzles revealed by observations in recent years is the apparent systematic growth of turbulence with redshift.
As traced by observations of bright nebular emission lines,
numerous studies
\citep[e.g.,][]{glazebrook13,simons17,johnson18,ubler19}
have confirmed a systematic increase 
of the velocity dispersion $\sigma_{\Ha}$
from $\sim 20$ \kms at $z=0$ \citep{terlevich81,epinat08,varidel16,zhou17} to $\sim 70$ \kms at $z \sim 2-3$ \citep[e.g.,][]{law09,fs09,fs18,jones10,price19}.  Indeed, in the $z \sim 2$ universe 
many typical star forming galaxies exhibit kinematics that are dominated by turbulent motion, even in cases for which
the galaxy morphology exhibits unambiguous spiral disk structure \citep{law12} \citep[although c.f.][]{yuan17}.

The cause of this strong evolution in turbulence is uncertain \citep[see review by][]{glazebrook13}
but has been broadly ascribed either to increased radiative and mechanical feedback from
intense star formation \citep[e.g.,][]{lehnert09, green14, moiseev15} or gravitational instabilities driven by galactic gas accretion
\citep[e.g.,][]{dekel09, ceverino10, krumholz16}, both of which predict a degree of correlation between 
$\sigma_{\Ha}$ and the total galactic star formation rate.
Distinguishing between these two scenarios is challenging;
in the high-redshift universe the entire star-forming main sequence
is offset to significantly higher star formation rates (SFR) at fixed stellar mass compared to $z=0$ \citep[see, e.g.,][]{wuyts11}, and at the same
time models also suggest the likelihood of significant gas accretion from clumpy cold streams \citep[e.g.,][]{dekel09, ceverino10}.  
In the nearby universe ultraluminous galaxies with high SFR possess much higher $\sigma_{\Ha}$ 
than their main sequence counterparts, yet localized peaks in the velocity dispersion
tend to be associated with large-scale 
gas flows rather than peaks in the star formation rate surface density \citep[e.g.,][]{colina05, arribas14}.

Generally, the large statistical studies of spectroscopic galaxy properties at $z \sim 0$ around which such studies are anchored are based on spatially unresolved fiber spectroscopy 
\citep[e.g.,][]{bman04}
for which $\sigma_{\Ha}$ can be both biased by contributions from active galactic nuclei (AGN) and 
challenging to separate from large-scale streaming and rotational motions.  Similarly, targeted spatially-resolved observations 
have by necessity focused on relatively small samples of objects \citep[e.g.,][]{andersen06, epinat10, martinsson13, arribas14, moiseev15, varidel16}
that trace only subsets of the star forming main sequence.

The early generation of IFU galaxy surveys has likewise not been able to address this question; the $R = 850$ spectral resolution
of the  Calar Alto Legacy Integral Field Area Survey \citep[CALIFA;][]{sanchez12} was too low to be able to measure \Ha\ velocity dispersions of typical 
galactic disks, while the $R \sim 1200$ Atlas3D survey \citep{cappellari11} focused exclusively on early-type galaxies.
With the current generation of large-scale IFU galaxy survey such as MaNGA \citep{bundy15} and SAMI \citep{croom12} , it has now become possible
for the first time to study statistically large representative samples of nearby galaxies with IFU observations similar to those employed at high redshifts.
The MaNGA survey
in particular represents more than an order of magnitude increase in sample size compared to previous IFU surveys,
with a now-completed sample of $>$ 10,000 galaxies
and contiguous spectral coverage in the wavelength range $\lambda\lambda 0.36 - 1.0 \micron$).  
However, the $R \sim 2000$ spectral resolution of MaNGA (corresponding to an instrumental resolution with $1\sigma$
width $\sim 70$ \kms) makes it challenging to study velocity dispersions that are typically on the order of
$20-30$ \kms.

In \citet[][]{paper1} we showed that through a multi-year concerted effort to understand and model the properties
of the MaNGA instrument, pipeline estimates of the instrumental line spread function (LSF) in the final
DR17 survey data products are
accurate to within 0.3\% systematic and 2\% random error, permitting 
the reliable study of astrophysical velocity dispersions down to 20 \kms and below.
Further, in \citet[][]{paper2} we showed that these astrophysical velocity dispersions exhibit dramatic trends with
the physical origin of the ionizing photons exciting the gas, with star-formation
activity preferentially occurring in dynamically cold disks with a well defined
peak in the galactic
line-of-sight velocity distribution around 25 \kms\ 
and AGN/LI(N)ER activity preferentially illuminating gas with much larger
velocity dispersions extending to 200 \kms\ and above.
In this third paper of the series, we focus on subtle variations in velocity dispersion within the star-forming
sequence and the underlying relation to the galactic star formation rate as a local benchmark for future observations at high redshift.

We structure our discussion as follows.
In \S \ref{obs.sec} we discuss the characteristics of the MaNGA galaxy sample and the relevant survey data products, and select a subset of spaxels whose line ratios are consistent with ionization by galactic HII regions.
In \S \ref{results.sec} we present the basic correlations between $\sigma_{\Ha}$ and star formation rate noting the effect of
beam smearing, and compare our results against recent works in the literature.
In \S \ref{details.sec} we use the statistical power of MaNGA to dissect the
contributions from star formation rate, stellar mass, and other observables in order to isolate the physical mechanism responsible
for the observed correlations.  We investigate the role of galaxy inclination in \S \ref{inclin.sec}, and in combination with galaxy azimuthal
information measure the ellipsoid of the ionized gas velocity dispersion field.
Finally, in \S \ref{discussion.sec} we discuss the implications of our results for the structure of galactic HII regions,
the underlying molecular gas, and for the diffuse ionized gas.
We summarize our conclusions in \S \ref{summary.sec}.

Throughout our analysis we adopt a \citet{chabrier03} stellar initial mass function and a 
$\Lambda$CDM cosmology in which $H_0 = 70$ km s$^{-1}$ Mpc$^{-1}$, $\Omega_m = 0.27$, and $\Omega_{\Lambda} = 0.73$.


\section{Observational Data}
\label{obs.sec}


\subsection{Survey Overview and Data Products}

The Mapping Nearby Galaxies at APO \citep[MaNGA,][]{bundy15} survey
is one of three primary surveys undertaken as part of the fourth-generation
Sloan Digital Sky Survey \citep[SDSS-IV,][]{blanton17} that uses a
multiplexed IFU fiber bundle interface \citep{drory15} to feed the
BOSS spectrographs \citep{smee13} on the Sloan 2.5m telescope \citep{gunn06}
at Apache Point Observatory.
MaNGA survey operations \citep{law15,yan16b} began in 2014 and concluded in 2020, providing
$R \sim 2000$ resolved spectroscopy in the wavelength range $\lambda\lambda 3600-10300$ \AA\ for $> 10,000$ unique galaxies
\citep[see][their Table 1]{paper1}.  These raw observational data
have been fully processed using the MaNGA Data Reduction Pipeline (DRP) to
produce science-grade calibrated data products \citep{law16,paper1,yan16a},
the final version of which was available internally to the SDSS-IV collaboration
as version MPL-11 and released to the broader astronomical community
as SDSS Data Release 17 \citep[DR17;][]{dr17} in December 2021.\footnote{DR17 is available both as flat FITS format data files and through the MARVIN python-based
framework \citep{cherinka19}; see https://www.sdss.org/dr17/}

Following \citet{paper1} and \citet{paper2}, we use the \Ha\ velocity dispersion maps produced
by the MaNGA Data Analysis Pipeline \citep[DAP,][]{belfiore19,westfall19}.
The stellar continuum templates used for fitting the emission lines were based on 
hierarchically clustered template spectra observed by the MaNGA MaStar program
\citep{yan19}.
These DAP velocity dispersion maps provide raw line widths, from which
we subtracted the expected instrumental LSF (also provided by the DAP) in quadrature.
We additionally corrected these maps for beam smearing arising from the finite
size of the MaNGA PSF by following the method described by \citet{paper1}
and \citet{paper2} (see their \S 5 and \S 2 respectively)
in which a model velocity field for each galaxy is
convolved with the PSF to estimate the artificial inflation
in the observed velocity dispersions.

As discussed at length by \citet{paper1} these DR17-based velocity dispersions
are significantly more reliable than the values provided in previous MaNGA public
data releases.


\subsection{MaNGA Galaxy Sample}

The full MaNGA galaxy sample is highly heterogenous and spans galaxies of a wide range of masses and morphological types with
a nearly flat mass distribution in the range $M_{\ast} = 10^9 - 10^{11} M_{\odot}$. 
This full sample is composed of a 
primary galaxy sample that covers a radial range out to 1.5 effective radii ($R_{\rm e}$), 
a secondary sample that covers out to 2.5 $R_{\rm e}$, a color-enhanced
sample that fills in less-represented regions of color/magnitude space, and a variety of ancillary programs that
have their own unique selection criteria
(e.g., massive galaxies, dwarf galaxies, Milky Way analogs, bright AGN, post starburst galaxies, etc).\footnote{Both stellar masses and effective radii are drawn from the parent galaxy catalog described
by \citet{wake17} based on an extension of the NASA-Sloan Atlas \citep[NSA;][]{blanton11}.}
In the present contribution we are most interested in the properties of star forming regions within star forming galaxies,
and must therefore downselect from the full sample.

Starting with the 11,273 galaxy data cubes in DR17, we follow \citet[][see their Section 2]{paper2} in rejecting 
nearby galaxies ($z < 0.001$), mis-centered galaxies, galaxies with data quality problems, and a small number of unusual
galaxies in the Coma cluster.  This leaves a sample of 10,016 data cubes corresponding to 9883 unique galaxies.
Next, we downselect to include only star-forming galaxies as identified by their mean \Ha\ equivalent width
measured within 1 effective radius ($R_{\rm e}$) of the galaxy center.
As illustrated in \citet[][see their Figure 1]{paper2}, our requirement of $EW_{\Ha} > 5$ \AA\ does a good job of selecting galaxies that lie on the star-forming main sequence, eliminating those in both the red sequence
($EW_{\Ha} < 2$ \AA) and the transitional `green valley' (2 \AA\ $< EW_{\Ha} < 5$ \AA)
and reducing our total sample to 5142 galaxy data cubes.

As discussed further in \S \ref{inclin.sec}, we make no specific cuts on galaxy disk inclination to the line of sight
in order to explore the impact of inclination on our results.


\subsection{Selecting Star-Forming Spaxels}
\label{bpt.sec}

While we have thus identified our {\it galaxy} sample, we must additionally restrict our study to those
spaxels within these galaxies whose
\Ha\ emission is dominated by ionizing photons arising from star-forming HII regions.  As we explored in depth
in \citet{paper2}, resolved \Ha\ velocity dispersions from the entire MaNGA sample ($\sim 3.6$ million spaxels across
7400 individual galaxies) exhibit a clear two-component kinematic population consisting of a dynamically cold 
gas disk (LOSVDs with a strong peak around $\sigma_{\Ha} = 24$ \kms) and an extended warm tail
(LOSVDs extending to $\sigma_{\Ha} > 100$ \kms).
These two populations are strongly segregated from each other by their strong-line nebular flux ratios
(e.g., \ntwo/\Ha, \stwo/\Ha , \othree/\Hb, and \oone/\Ha) the boundary between which is traced by a series
of well defined curves \citep[see][their Eqns. 1-3]{paper2}.
For the \stwo/\Ha\ vs \othree/\Hb\ relation in particular, these kinematically defined curves are in excellent agreement
with theoretical relations \citep[e.g.,][]{kewley01}, indicating that selecting spaxels according to
their strong emission line ratios is a reliable way of identifying dynamically cold
star-forming gas disks.\footnote{Although
see \citet{paper2} for discussion of second-order effects and an old, dynamically warm tail of the LI(N)ER sequence that extends
throughout the traditional star-forming region of the \stwo/\Ha\ vs \othree/\Hb\ diagram.}

We therefore make our initial identification of star-forming spaxels by selecting those whose 
\stwo/\Ha\ vs \othree/\Hb\ line ratios place them below the relation defined by
\citet[][their Eqn. 2]{paper2}:

\begin{equation}
    R3 = \frac{0.648}{S2 - 0.324} + 1.349
    \label{eqn:R3}
\end{equation}
for $R3 \equiv \textrm{log}(\othree\ \lambda5007/\Hb)$ and $S2 \equiv \textrm{log}(\stwo\ \lambda6716 + \lambda6731/\Ha)$.
As discussed by \citet[][see their \S 5.4 and Figure 9]{paper2} this selection cut is broadly similar to the
criterion defined by \citet{kewley01}, except slightly offset to larger
$S2$ in order to compensate for star-forming spaxels at large galactocentric radii that were missing from
early work using SDSS single-fiber spectra against which the \citet{kewley01} study was calibrated.

As our goal is to diagnose statistical trends in actively star-forming galaxies for which $\sigma_{\Ha}$ is
typically $\leq 30$ \kms, we additionally restrict our sample to spaxels with \Ha\ signal-to-noise ratio (SNR)
$> 50$ and that lie at radii $> 4$ arcsec from the center of each galaxy.\footnote{For comparison, the MaNGA galaxy fiber bundles range from 6 to 16 arcsec in radius; see 
\citep[][their Figure 1]{wake17} for the corresponding distribution of values for $R_e$.}  The first of these cuts
is necessary in order to avoid systematic bias in the recovered velocity dispersions arising from the preferential loss of spaxels whose measured line widths scatter below the instrumental resolution at
low SNR \citep[see discussion in][their \S 4.3 and Figure 15]{paper1}.  The second of these cuts allows us to mitigate additional systematic biases that arise from beam smearing by the observational PSF, which is most pronounced in the central regions of galaxies.  Although we have corrected the DAP velocity dispersion maps
for beam smearing following the method outlined by \citet[][their \S 5]{paper1}, as these authors note the
corrected values are most uncertain within the central 4 arcsec.
Additionally, we require the \stwo, \othree, and \Hb\ lines to be detected at SNR $> 3$ to avoid
contamination by low-SNR line ratios, and $\sigma_{\Ha} < 100$ \kms.  This final kinematic selection
serves only to eliminate $\sim 3000$ abnormally-high $\sigma_{\Ha}$ spaxels from the sample; these are predominantly from galaxies that have clear major-merger morphology and multi-component LOSVDs that the DAP was not designed to fit reliably.

\begin{figure}[!]
\epsscale{1.2}
\plotone{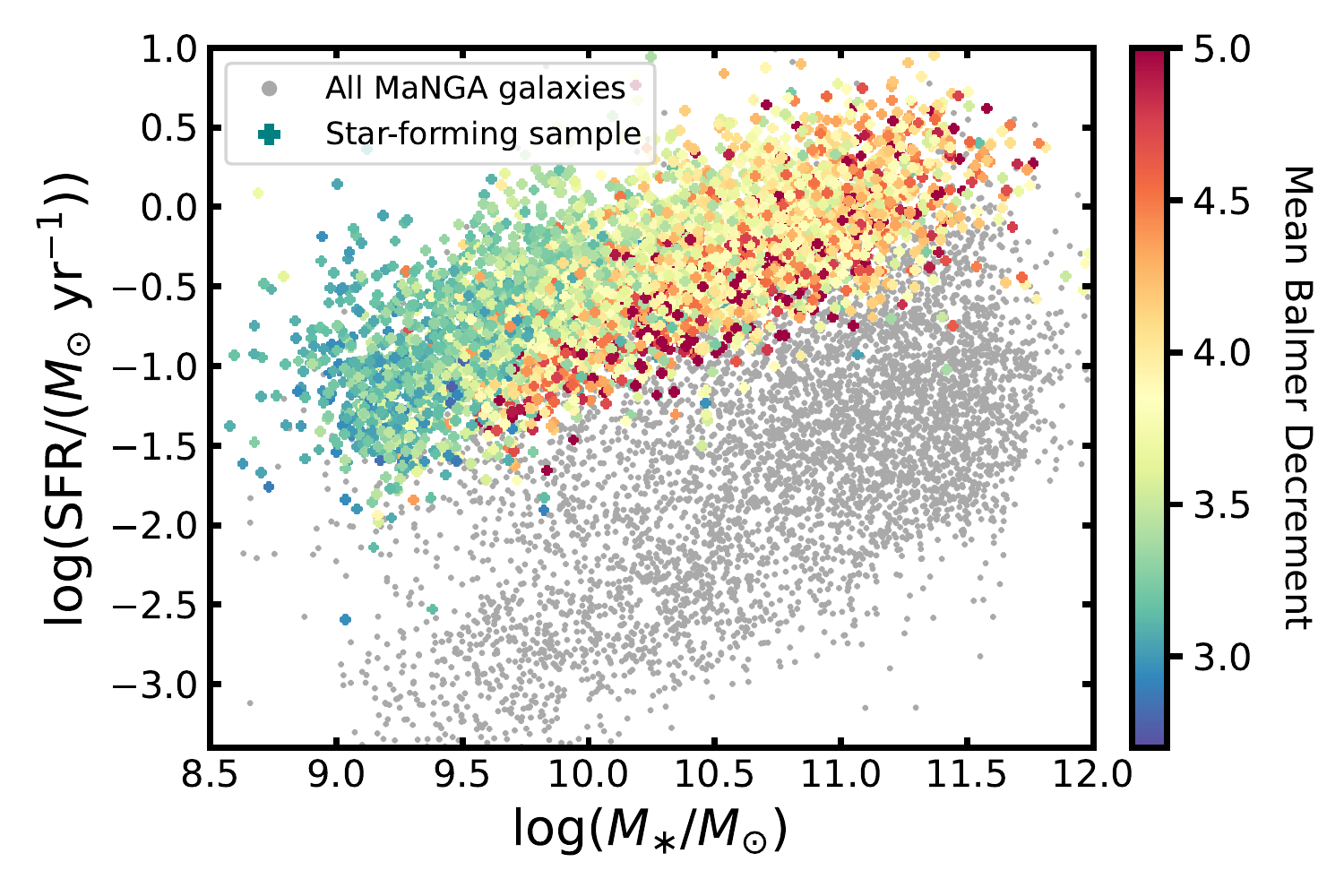}
\caption{Total SFR (computed from the summation of all MaNGA spaxels, without dust
correction) as a function of stellar mass for the entire MaNGA DR17 galaxy sample
(grey points).  The 4517 star-forming galaxies used for our 
analysis are color-coded by the mean Balmer decrement ($f_{\Ha}/f_{\Hb}$) of their
star-forming spaxels meeting our selection criteria \citep[c.f. Figure 1 of][]{paper2}.
}
\label{sample.fig}
\end{figure}

After all of these cuts, our final sample size is 1.4 million spaxels\footnote{As discussed in \citet{paper2}, the number of statistically independent spectra will be somewhat lower than this ($\sim 100,000 - 500,000$) given correlations on the angular scale of the MaNGA PSF.} from 4,517 individual galaxies.  We show the distribution of these galaxies
compared to the full MaNGA sample in Figure \ref{sample.fig}.

\section{Results}
\label{results.sec}


\subsection{The Local Relation}
\label{local.sec}

Following the generic predictions of star formation feedback models \citep[e.g.,][and references therein]{krumholz16, hayward17, hung19} and results from a variety of
observational studies \citep[e.g.,][]{green14, varidel20},
we expect that the MaNGA data should exhibit 
a positive correlation between the local gas-phase velocity
dispersion in each spaxel ($\sigma_{\Ha}$) and the corresponding local
star formation rate surface density ($\Sigma_{\rm SFR}$).

\begin{figure*}[!]
\epsscale{1.1}
\plotone{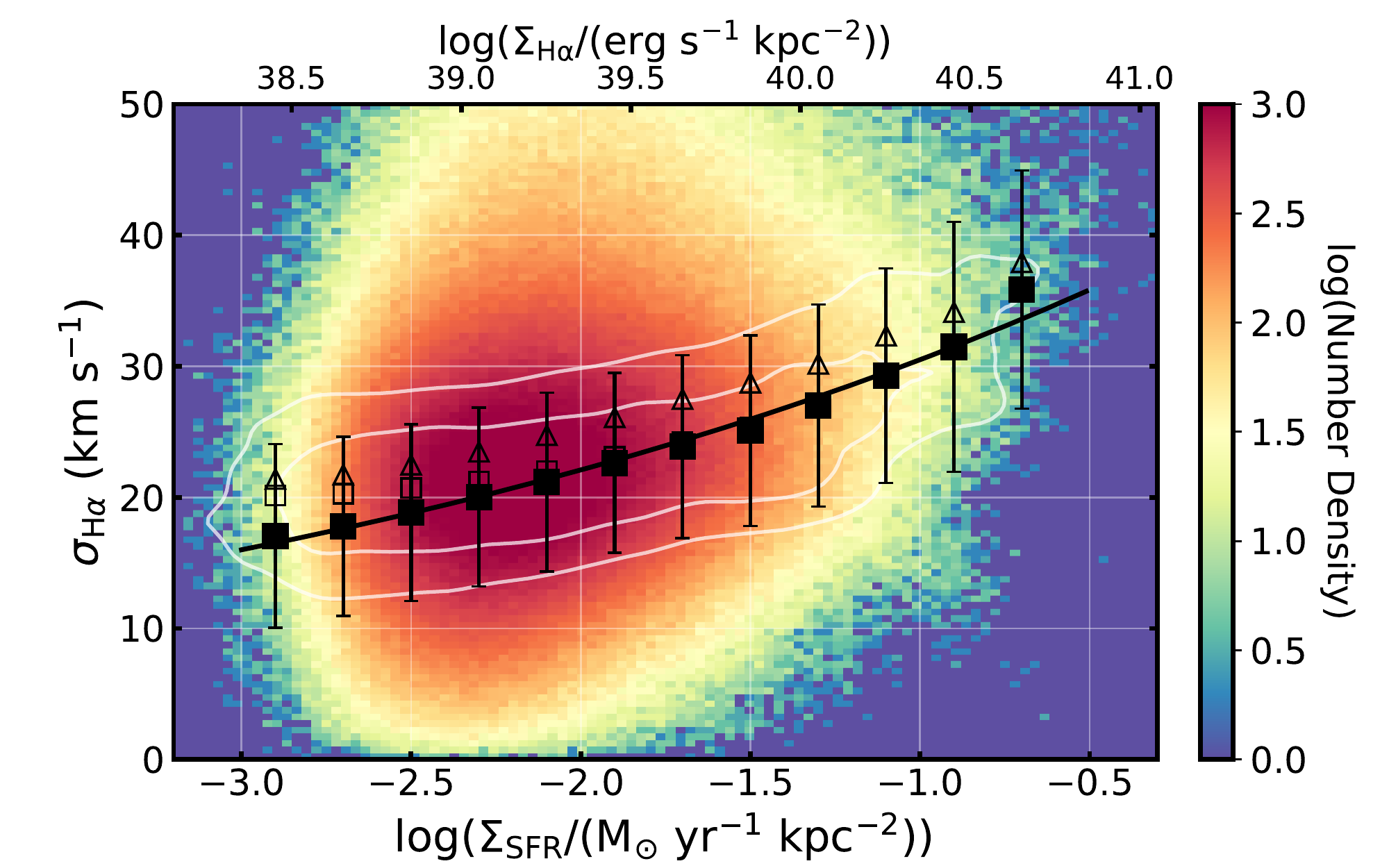}
\caption{Logarithmic number density plot of local
\Ha\ velocity dispersion as a function of dust-corrected star formation rate surface density and \Ha\ surface brightness
for 1.4 million spaxels from 4,517 individual galaxies in the MaNGA DR17
final data release.  Grey contours represent 50\% and 75\% intervals in the 
number density distribution when counts are normalized in each bin
along the $\Sigma_{\rm SFR}$ axis.
Open symbols represent the moving sigma-clipped average of the distribution with
(open boxes) and without (open triangles) a beam smearing correction applied to the
spaxel data.  Black filled squares are as the open boxes, but with an additional
correction applied to account for the preferential loss of values at low $\sigma_{\Ha}$
from the distribution due to the MaNGA LSF (see discussion in text).  Error bars on
the black filled squares represent the $1\sigma$ width of
the distribution corrected in quadrature for the expected width due to the observational uncertainties in $\sigma_{\Ha}$ and the MaNGA LSF.  The black line represents a power-law fit to the relation given by Eqn. \ref{local.eqn}.
}
\label{sigma_sfrsd.fig}
\end{figure*}

In Figure \ref{sigma_sfrsd.fig} we therefore plot $\sigma_{\Ha}$ against
$\Sigma_{\rm SFR}$ for all 1.4 million spaxels in our DR17 star-forming sample,
color-coded by the logarithm of the number of spaxels contributing to each 
location in the plot.
In all cases $\sigma_{\Ha}$ for each spaxel is drawn directly from the DAP-provided data
products and corrected for the instrumental LSF and a model-derived estimate
of the beam smearing imparted by the MaNGA PSF \citep[see][their \S 5]{paper1}.
$\Sigma_{\rm SFR}$ is similarly derived from the MaNGA DAP data products by converting the reported \Ha\ flux for each spaxel to the
\Ha\ SFR following the relation defined by \citet[][their Eqn 2]{kennicutt98}:
\begin{equation}
\Sigma_{\rm SFR} (M_{\odot} \textrm{yr}^{-1} \textrm{kpc}^{-2}) = \frac{L(\Ha)}{1.26 \times 10^{41} \textrm{erg s}^{-1}} \times (0.56/\Theta)
\label{eqn:SSFR}
\end{equation}
where $L_{\Ha}$ in units of erg s$^{-1}$ is the dust-corrected
\Ha\ luminosity of the spaxel given the known redshift
of the source, the factor of 0.56 represents a conversion to the \citet{chabrier03} initial mass function from that used by \citet{kennicutt98}, and $\Theta$ is a scale factor representing the solid angle of a spaxel in square kpc.  Our dust correction factor for
each spaxel is based upon the corresponding MaNGA-observed
\Ha/\Hb\ Balmer decrement (assuming an unreddened ratio \Ha/\Hb\ $= 2.86$)
in combination with a \citet{cardelli89} dust model.

Although there is substantial scatter in the measured velocity dispersion
of individual spaxels \citep[due primarily to the 2\% random uncertainty in the MaNGA
LSF around \Ha,][]{paper1}, there is a clear overall trend in the sense that
gas-phase velocity dispersion increases across the range
of star-formation rate surface densities probed by MaNGA.  This trend is driven
largely by the absence of spaxels at low $\sigma_{\Ha}$ and high
$\Sigma_{\rm SFR}$; this absence is likely to be genuine rather than a selection
bias as narrow, bright emission lines should be particularly easy to detect.

It is nonetheless difficult to get a sense of the overall trends in star-forming galaxies as an ensemble from this plot as 
the regions of low spaxel density confuse the impression of
the statistical bulk of the spaxels which are
concentrated in a narrow range around log($\Sigma_{\rm SFR}$/($M_{\odot} {\rm yr}^{-1} {\rm kpc}^{-2}$)) = -2.25 and $\sigma_{\Ha} = 20-25$ \kms\ \citep[see also Figure 3 of ][]{paper2}.  We therefore compute the $2.5\sigma$-clipped mean\footnote{Our mean $\sigma_{\Ha}$ changes
by about 0.5 \kms\ if we use a less-restrictive $5\sigma$ cut, or simply compute
the median value in each bin.}
of the distribution in bins of $\Sigma_{\rm SFR}$, overplotting this running mean
against the raw data in open black squares in Figure \ref{sigma_sfrsd.fig}.
This spaxel-averaged relation shows a clear trend, but flattens out substantially
at the lowest values of $\sigma_{\Ha}$ due to the survival bias imparted by the large
MaNGA LSF.  That is, lower values of $\sigma_{\Ha}$ are preferentially lost
from the sample as observational uncertainties scatter them below the instrumental LSF,
resulting in imaginary velocity dispersions after correction for the LSF in quadrature.
We explored this effect in detail in \citet{paper1} and provided there
a series of correction factors to account for this effect as a function of both SNR
and the intrinsic astrophysical velocity dispersion.  Per their Figure 15, this
correction can be as large as 10\% at $\sigma_{\Ha} = 15$ \kms, even for SNR=100.

\begin{deluxetable*}{ccccccccccc}
\label{results.table}
\tablecolumns{11}
\tablewidth{0pc}
\tabletypesize{\scriptsize}
\tablecaption{MaNGA Ensemble Velocity Dispersions}
\tablehead{
\multicolumn{3}{c}{SFR Surface Density} & \colhead{} & \multicolumn{3}{c}{SFR} & \colhead{} & \multicolumn{3}{c}{SFR (1$R_{\rm e}$)} \\
\cline{1-3}  \cline{5-7} \cline{9-11}\\
\colhead{log($\Sigma_{\rm SFR}$)} & \colhead{$\sigma_{\Ha}$} & \colhead{$\epsilon_{\Ha}$} & \colhead{} &
\colhead{log(SFR)} & \colhead{$\langle\sigma_{\Ha}\rangle$} & \colhead{$\epsilon_{\Ha}$} & \colhead{} & 
\colhead{log(SFR)} & \colhead{$\langle\sigma_{\Ha}\rangle$} & \colhead{$\epsilon_{\Ha}$} \\
\colhead{($M_{\odot}$ yr$^{-1}$ kpc$^{-2}$)} & \colhead{(\kms)} & \colhead{(\kms)} & \colhead{} &
\colhead{($M_{\odot}$ yr$^{-1}$)} & \colhead{(\kms)} & \colhead{(\kms)} & \colhead{} & 
\colhead{($M_{\odot}$ yr$^{-1}$)} & \colhead{(\kms)} & \colhead{(\kms)} \\
}
\startdata
-2.9 & 17.1 & 7.0 &  & -1.50 & 16.1 & 4.1 &  & -1.70 & 15.5 & 4.2\\
-2.7 & 17.7 & 6.8 &  & -1.28 & 15.5 & 3.1 &  & -1.48 & 16.4 & 3.8\\
-2.5 & 18.8 & 6.7 &  & -1.06 & 16.8 & 3.3 &  & -1.26 & 17.0 & 3.5\\
-2.3 & 20.0 & 6.8 &  & -0.84 & 18.3 & 4.0 &  & -1.04 & 18.4 & 4.0\\
-2.1 & 21.2 & 6.8 &  & -0.62 & 19.0 & 3.8 &  & -0.82 & 19.1 & 3.8\\
-1.9 & 22.6 & 6.9 &  & -0.40 & 20.1 & 4.0 &  & -0.60 & 20.5 & 4.4\\
-1.7 & 23.9 & 7.0 &  & -0.18 & 22.2 & 5.3 &  & -0.38 & 22.2 & 5.3\\
-1.5 & 25.1 & 7.3 &  &  0.04 & 23.6 & 5.7 &  & -0.16 & 23.6 & 5.2\\
-1.3 & 27.0 & 7.7 &  &  0.26 & 24.8 & 5.3 &  &  0.06 & 25.7 & 5.7\\
-1.1 & 29.3 & 8.2 &  &  0.48 & 26.7 & 5.7 &  &  0.28 & 28.4 & 6.7\\
-0.9 & 31.5 & 9.5 &  &  0.70 & 29.8 & 6.3 &  &  0.50 & 30.0 & 6.7\\
-0.7 & 35.8 & 9.1 &  &  0.92 & 33.0 & 7.2 &  &  0.72 & 33.5 & 7.2
\enddata
\end{deluxetable*}

We therefore apply a survival bias correction to the open black squares following
the prescriptions of \citet{paper1} in order to obtain the relation given by the filled
black squares in Figure \ref{sigma_sfrsd.fig} 
(see also Table \ref{results.table})
with error bars representing the rms width of the distribution corrected for
observational uncertainties.\footnote{I.e., the measured rms width of the
distribution with the contribution from observational uncertainty in $\sigma_{\Ha}$
subtracted in quadrature.  The observational uncertainty in $\sigma_{\Ha}$ is
computed as the measured uncertainty in the line width (provided by the DAP) combined with a 2\% stochastic uncertainty in the MaNGA LSF \citep[see comparisons against extensive
Monte Carlo simulations given by][their Figure 15]{paper1}.}
This corrected relation is much clearer, showing that 
$\sigma_{\Ha}$ increases from 17 \kms\ to 35 \kms\ over about two decades in
$\Sigma_{\rm SFR}$ from $10^{-2.9}$ to $10^{-0.7}$ $M_{\odot}$ yr$^{-1}$ kpc$^{-2}$.
Using an unweighted least-squares fitting algorithm in log-log space, we find that this corresponds
to an approximately power-law relation of the form
\begin{equation}
\sigma_{\Ha} = 42 \pm 1 \left(\frac{\Sigma_{\rm SFR}}{M_{\odot} \, {\rm yr}^{-1} \,  {\rm kpc}^{-2}}\right)^{0.140 \pm 0.005} \rm{km \, s}^{-1}
\label{local.eqn}
\end{equation}

Despite our focus on on spaxels outside the central 4 arcsec in each galaxy, we
nonetheless note that the strength of this correlation depends at the
few \kms\ level on our beam smearing correction.  If we repeat the above analysis
without making such a correction, we derive values that are systematically larger
by 1 \kms\ at the low end of the relation and 3 \kms\ at the high end of the relation
(open triangles vs open boxes in Figure \ref{sigma_sfrsd.fig}).
We return to a discussion of this effect in \S \ref{details.sec}.

\subsection{The Global Relation}
\label{global.sec}

In addition to presenting our results in a local sense for individual spaxels as in \S \ref{local.sec},
it is also possible to recast our observations in terms of galaxy-averaged
quantities.
This approach has, for instance, been commonly adopted
in studies of the high-redshift universe \citep[e.g.,][]{law09,wisnioski15,simons17,ubler19} for which a single 
velocity dispersion is typically quoted for each galaxy.
The method of estimating that single velocity dispersion value
for each galaxy can vary significantly from study to study,
with some preferring to average individual measurements and others
constructing physically-based models of the galaxies that can
be matched to the ensemble of observed spaxels via forward modeling.
Here, we follow \citet{law09} and \citet{green14} 
in computing the intensity-weighted
mean velocity dispersion $\langle \sigma_{\Ha} \rangle$
of spaxels in each galaxy that fulfill
the selection criteria defined in \S \ref{bpt.sec}.

\begin{figure*}[!]
\epsscale{1.1}
\plotone{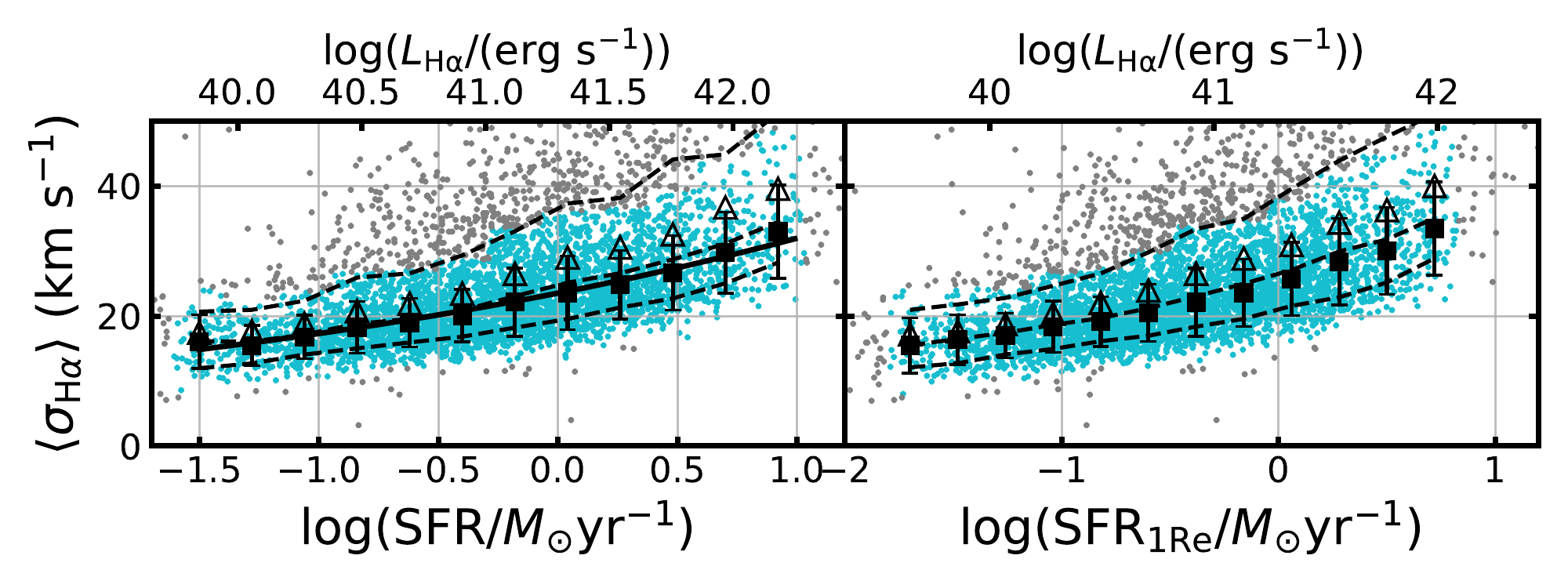}
\caption{Intensity averaged \Ha\ velocity dispersion as a function of the total
star formation rate in the MaNGA footprint (left panel) and within one effective
elliptical radius of the galaxy center (right panel) for 4517 star-forming galaxies
from MaNGA DR17 (small cyan and grey points).  Black filled squares represent a moving sigma-clipped average
with error bars representing the sigma-clipped rms distribution
width.  Grey/cyan points represent data rejected/kept by the clipping algorithm.  Black open triangles represent the same quantity as the black filled squares but derived from velocity dispersions that have not been corrected for beam smearing.  The solid black line represents a power-law fit to the relation given by Eqn. \ref{global.eqn}.  Dashed black lines represent the 16th, 50th, and 84th percentiles of the full distribution.
}
\label{paper3_global.fig}
\end{figure*}

We plot this galaxy-averaged velocity dispersion in Figure \ref{paper3_global.fig} against two estimates of the 
total dust-corrected star formation rate; the total star formation rate within the MaNGA IFU footprint, and the total star formation
rate within 1 effective radius of the galaxy center.  The former value is computed by simply summing the
\Ha-derived star formation rates within each galaxy data cube for all spaxels in which the nebular emission line 
ratios are consistent with ionization from star formation (\S \ref{bpt.sec}), while the latter includes only spaxels at radii
less than one $R_{\rm e}$ (given in elliptical polar coordinates by the MaNGA DAP for each
galaxy).\footnote{If we recompute the galaxy-averaged velocity dispersion using only spaxels at $r < R_{\rm e}$ for consistency, $\sim$ 700 fewer galaxies populate the right-hand panel of Figure \ref{paper3_global.fig} but the averages (filled black squares) change by just 0.4 \kms.  Likewise, the coefficients in Equation \ref{global.eqn} change by amounts less than or comparable to the quoated $1\sigma$ uncertainty.}
While the simple sum across the MaNGA footprint will be a closer approximation to the total
galaxy SFR, the SFR within $1R_{\rm e}$ will be a more uniform statistic across the MaNGA sample, as individual galaxies
can be covered out to a range of different radii along their major and minor axes.

As before, we also compute the $2.5\sigma$-clipped mean
of the distribution in bins of total SFR\footnote{This sigma-clipped mean matches the 50th percentile of the overall distribution to within
$< 1$ \kms\ on average.}, overplotting this running mean
against the raw data in filled black squares in Figure \ref{paper3_global.fig} (see also Table \ref{results.table})
with error bars representing the rms width of the clipped distribution.
No attempt has been made in this estimate
to remove the observational uncertainty in individual line measurements, as such uncertainties largely average to zero
in our construction of $\langle \sigma_{\Ha} \rangle$ from the individual spaxel measurements.  Likewise, survival bias in the individual spaxel measurements is largely
mitigated by the intensity weighting within each galaxy.
For both the total SFR within the MaNGA footprint (left panel) and the total SFR within $1R_{\rm e}$ (right panel), we find
that the local correlation between $\sigma_{\Ha}$
and $\Sigma_{\rm SFR}$ extends in a global sense to a 
strong correlation between $\langle \sigma_{\Ha} \rangle$
and the total galaxy-integrated SFR as well, increasing from 16 \kms\ to 33 \kms\ over 2.5 decades in total SFR.
This relation is again well described by a power-law relation, given by 
\begin{equation}
\langle \sigma_{\Ha} \rangle = 23.6 \pm 0.3 \left(\frac{{\rm SFR}}{M_{\odot} \, {\rm yr}^{-1}}\right)^{0.133 \pm 0.006} \rm{km \, s}^{-1}
\label{global.eqn}
\end{equation}
As for the local relation, uncorrected beam smearing can exaggerate the strength 
of this correlation (open triangles in Figure \ref{paper3_global.fig}) but
does not wholly explain it.

We note that Figure \ref{paper3_global.fig} contains an appreciable number of galaxies
for which $\langle \sigma_{\Ha} \rangle$ lies above the general relation with
the tail of the distribution extending above the plotted range to
$\sim$ 120 \kms.  Of the 4517 galaxies in this plot, 665 are effectively removed
from consideration in deriving the average relation
by our $2.5\sigma$ clipping algorithm: 135 have $\langle \sigma_{\Ha} \rangle > 50$ \kms,
and 7 have $\langle \sigma_{\Ha} \rangle > 100$ \kms.

The physical origin of this high-dispersion tail varies significantly from galaxy to galaxy.  Galaxies with 
$\langle \sigma_{\Ha} \rangle > 50$ \kms\
are $\sim 10^{\circ}$ more highly inclined to the line of sight on average than
the rest of the galaxy sample, suggesting that the line-of-sight velocity dispersion 
in some cases may be inflated by contributions from the rotational velocity field.
In a handful of galaxies randomly selected for inspection from the high-$\langle \sigma_{\Ha} \rangle$ tail however
(e.g., 7977-12701, 8332-12704) visual inspection of the galaxy spectra 
indicates clear 
multi-component nebular emission lines indicative of bulk
gas flows that cannot be well fit by the single-component gaussian models used by the DAP.
We defer treatment of such multi-component profiles to future work, and simply note here that they
represent a small fraction of the overall MaNGA galaxy sample.


\subsection{Comparison to Recent Literature}

\begin{figure*}[!]
\epsscale{1.2}
\plotone{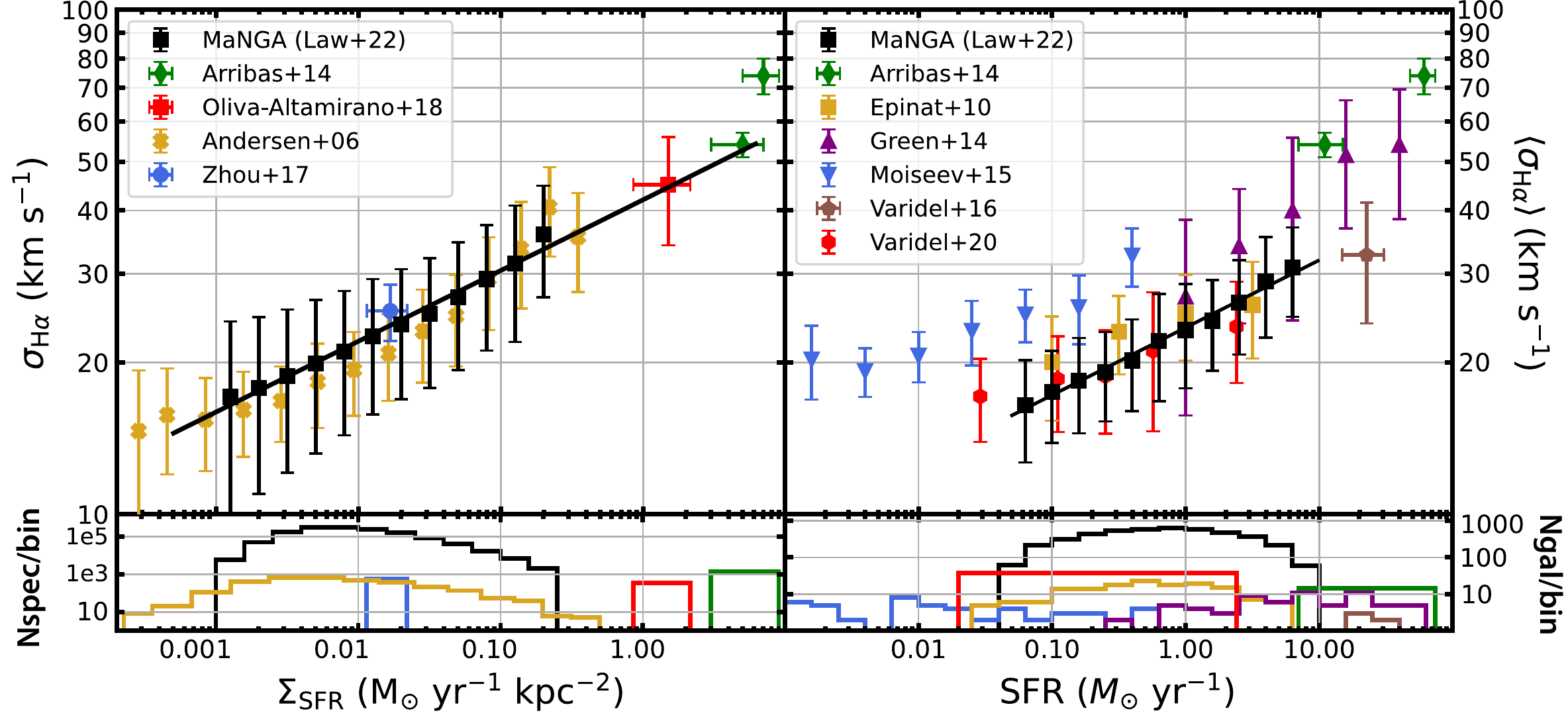}
\caption{Per-spaxel velocity dispersion $\sigma_{\Ha}$ as a function of the
local star formation rate surface density $\Sigma_{\rm SFR}$ (left panel)
and galaxy-averaged $\langle \sigma_{\Ha} \rangle$ as a function of the galaxy-integrated
star formation rate (right panel) for MaNGA DR17 galaxies
and other $z=0$ samples from the literature.  
Star formation rates for all samples are consistently shown dust-corrected using a \citet{chabrier03} IMF.
As in Figures \ref{sigma_sfrsd.fig} and \ref{paper3_global.fig} the MaNGA
data points represent moving averages computed for 4517 MaNGA galaxies
with error bars showing the clipped
$1\sigma$ range of the distribution after correction
for the additional scatter introduced by the uncertainty in the LSF.
Literature values are similarly averaged across individual galaxies with error
bars showing the range of the distribution, including 32/16 LIRGs/ULIRGs
from \citet{arribas14}, 7 star forming galaxies from \citet{zhou17},
7 DYNAMO galaxies with 400 pc adaptive-optics resolution observations
from \citet{oa18},
59 dwarf galaxies from \citet{moiseev15}, 39 spiral galaxies from
\citet{andersen06}, 153 galaxies from \citet{epinat10}, six starburst galaxies
from \citet{varidel16},  67 starburst galaxies from \citet{green14}, and 383 galaxies from the combined SAMI+DYNAMO
sample \citep{varidel20}.  The histograms in the lower panels show the logarithmic number of spectra (left-hand panel) or galaxies (right-hand panel) contributing
to each bin in SFR or $\Sigma_{\rm SFR}$ for each of the different samples (assuming 100 spectra per galaxy for the \citet{arribas14}, \citet{oa18}, and \citet{zhou17} samples).
}
\label{litcompare.fig}
\end{figure*}

While the MaNGA relation between $\sigma_{\Ha}$ and star formation rate 
shown in Figures \ref{sigma_sfrsd.fig} and \ref{paper3_global.fig} represents by far the largest
sample of galaxies from across the entire star forming main sequence to date, our results largely confirm previous
measurements made from smaller galaxy samples (often made at much higher spectral resolution).
Forty years ago for instance, 
\citet{terlevich81} measured gas-phase velocity dispersions for a selection of extragalactic HII regions,
finding typical $\sigma_{\Ha} = 15-30$ \kms\ in excellent agreement with the values that we have derived
from the overall star forming galaxy population.
In Figure \ref{litcompare.fig} we compare our observed MaNGA relations between $\sigma_{\Ha}$, the local star formation rate
surface density, and the total galactic star formation rate against a variety of other studies
from the recent literature.

\begin{figure}[!]
\epsscale{1.2}
\plotone{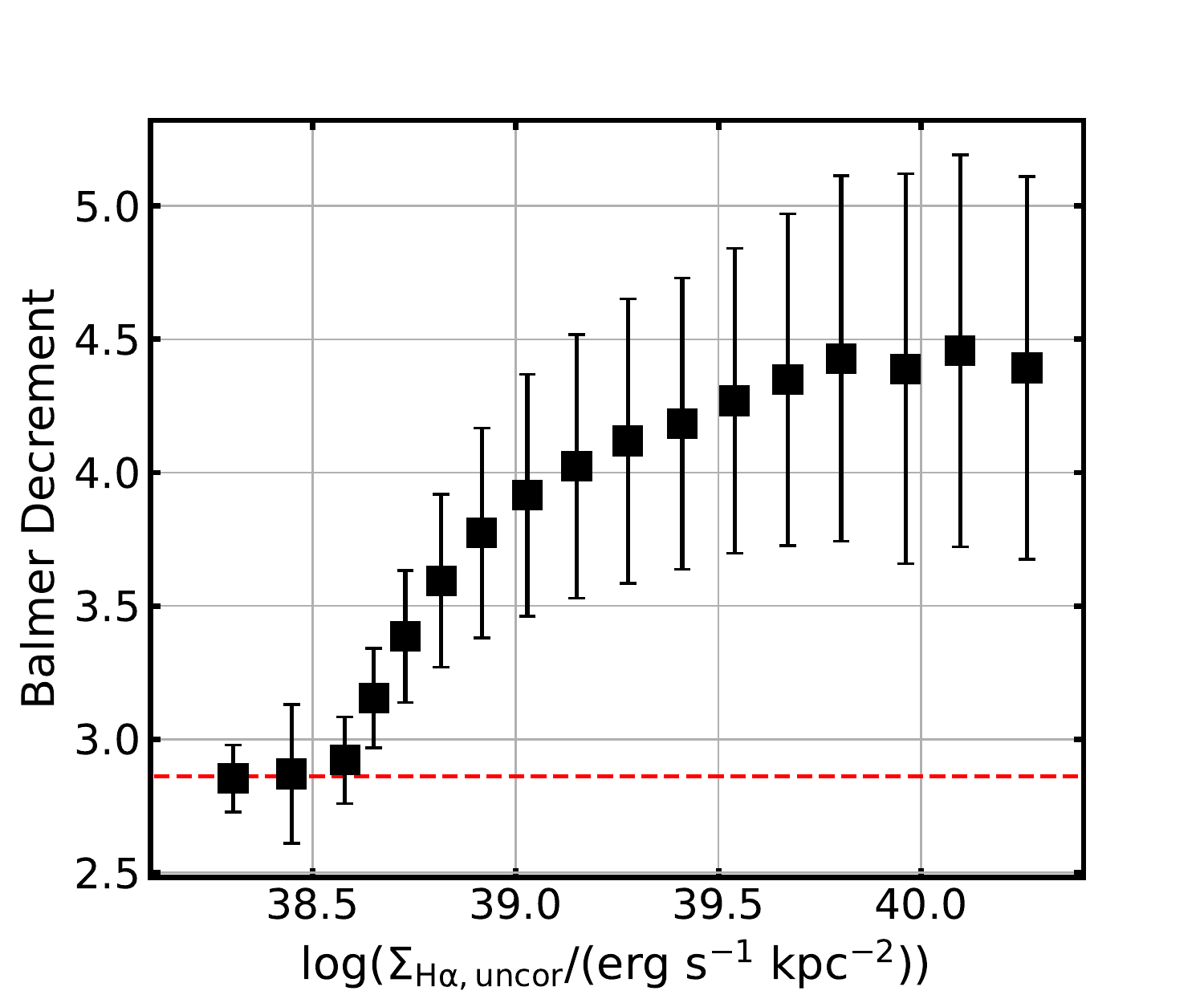}
\caption{Average Balmer decrement (i.e., \Ha/\Hb) measured from the MaNGA
spaxel data in Figure \ref{sigma_sfrsd.fig} as a function of the uncorrected \Ha\ surface brightness $\Sigma_{\rm \Ha, uncor}$.  Error bars indicate the $1\sigma$ width of the spaxel distribution.
The dashed red line represents the constant value of 2.86 corresponding to the nominal dust-free value under Case B recombination \citep{osterbrock06}.
}
\label{bdec.fig}
\end{figure}

\citet{andersen06} for instance found an average $\sigma_{\Ha} = 18 \pm 4$  \kms\ (see their Figure 6) from DensePak IFU \citep{barden98} echelle spectroscopy of 39 face-on spirals (mean inclination 23$^{\circ}$) with a comparable spatial resolution to MaNGA ($\sim 1.1$ kpc median fiber diameter). This agrees well with the low-$\Sigma_{\rm SFR}$ end of the MaNGA distribution where the peak number of spaxels are located for both the MaNGA and the DensePak data (Figure \ref{litcompare.fig}, lower panels). Reprocessing the DensePak line-width data from \citet{andersen06} with the flux calibration from \citet{andersen13} to present the results as a function of $\Sigma_{\rm SFR}$ (Figure \ref{litcompare.fig}, gold points in left-hand plot) we find excellent agreement in the trend of increasing $\sigma_{\Ha}$ with increasing $\Sigma_{\rm SFR}$ as well. We select only DensePak spectra consistent with star-forming regions using $\rm \log[NII]\lambda 6584/H\alpha < -0.3$ based on [NII] line-fluxes from D. Andersen (private comm.); this is a conservative selection in the context of Equation~\ref{eqn:R3}. We have applied Equation~\ref{eqn:SSFR} with a variable dust correction factor
assuming that the Balmer decrement varies from 2.86 to 4.5 as a function of
the \Ha\ surface brightness in a manner similar to that observed in the MaNGA
spaxel sample (see Figure \ref{bdec.fig}).
While no correction for beam-smearing has been applied, the line-widths have been corrected for the instrumental broadening. The DensePak and MaNGA relations are in excellent agreement to within
observational uncertainty, despite the DensePak data having substantially higher spectral resolution $R \sim 13,000$ providing an instrumental $1\sigma$ LSF of just 10 \kms.

Similarly, \citet{epinat08,epinat10} presented Fabry-Perot \Ha\ kinematics for a sample of $\sim 100$ nearby spiral galaxies
from the GHASP survey 
and found average \Ha\ velocity dispersion
$\langle \sigma_{\Ha} \rangle = 24 \pm 5$ \kms.  Binning the results from their online supplemental data
according to SFR (and applying a dust correction factor of 2.0 based on the average correction for the MaNGA star forming
galaxy sample), we find a relation that matches the MaNGA data to within an average of 1.2 \kms.

More recently, observations from the SAMI IFU survey have also been used to assess the velocity dispersion of galactic disks.
\citet{zhou17} for instance analyzed SAMI velocity maps for 8 local star 
forming galaxies\footnote{We ignore 
one of their eight sources with large uncertainties; see discussion by \citet{varidel20}.}
and found $\sigma_{\Ha}$  in the range 20-30 \kms\ for galaxies with $\Sigma_{\rm SFR} \approx 10^{-2} M_{\odot}$ yr$^{-1}$ kpc$^{-2}$,
in excellent agreement with our results (Figure \ref{litcompare.fig}, left-hand panel).
Likewise,  \citet{varidel20} analyzed a sample of 383 galaxies from the combined SAMI and DYNAMO samples
and found a statistically significant correlation between SFR and $\langle \sigma_{\Ha} \rangle$.
As indicated by Figure \ref{litcompare.fig} (right-hand panel), in the SFR range of overlap between the MaNGA and SAMI samples
the MaNGA velocity dispersions are larger on average by about 0.4 \kms, consistent
with the 0.7 \kms\ that we found in \citet[][see their Figure 21]{paper1} for a sample of galaxies
observed in common between the two surveys.
We note, however, that \citet{varidel20} made an inclination-dependent 
correction to their observations and thus estimate the vertical disk velocity dispersion
$\sigma_{\rm z}$
rather than the simple line-of-sight velocity dispersion.
As discussed in \S \ref{inclin.sec}, if we were to make such a correction our results
would shift downward by $\sim 1.5$ \kms, and nonetheless still be a reasonable match
to within 1.1 \kms.
Broadly speaking, the MaNGA and SAMI data thus give much the same trend, with the
higher spectral resolution SAMI data complemented by the 10x larger
MaNGA sample.

At lower star formation rates, 
\citet{moiseev15} used scanning Fabry-Perot interferometry to study the ionized gas in 59 nearby dwarf galaxies
(SFR 0.001 to 0.1 $M_{\odot}$ yr$^{-1}$).
While these authors observed a trend between $\langle \sigma_{\Ha} \rangle$ 
and SFR similar to ours, this trend appears offset
relative to the MaNGA, GHASP, and SAMI observations.
Adding a 3 \kms\ natural line width and 9 \kms\ thermal broadening back into their data in quadrature for consistency with
our observations (since these authors subtracted these quantities from their published
values), we note that
their 20-30 \kms\ values are systematically about 10 \kms\ higher than the MaNGA observations at similar SFR.\footnote{A similar discrepancy between the \citet{moiseev15}
results and the SAMI survey results was previously noted by \citet{varidel20}.}
The reason for this discrepancy is unclear; while it may reflect a genuine difference between the dwarf galaxy population
and the star forming main sequence at higher stellar masses, it may also be due in part to systematic differences between
the survey analysis techniques.

At higher star formation rates, 
\citet{green14} \citep[see also][]{green10} 
presented initial results from the DYNAMO survey of 67 $z \sim 0.1$ starburst galaxies (SFR $0.2 -57 M_{\odot}$/yr).
While these authors found a similar correlation between $\sigma_{\Ha}$ and total SFR, their relation is
steeper than ours, reaching $40 - 50$ \kms at $10 M_{\odot}$ yr$^{-1}$ instead of 30 \kms.
 \citet{arribas14} and \citet{gm09} likewise observed large values of $\sigma_{\Ha}$ for a sample of 
58 LIRGs and ULIRGs, as did
\citet{varidel16} for six $z < 0.04$ starburst galaxies observed with the WiFeS IFU, and
\citet{oa18} for Keck/OSIRIS IFU 
observations (at 400 pc adaptive-optics resolution) for a sample of 
7 disk galaxies reobserved from the DYNAMO sample.
As illustrated in Figure \ref{litcompare.fig} (left-hand panel), the \citet{arribas14} and \citet{oa18} samples
represent SFR surface densities 1-1.5 orders of magnitude larger than those probed by the MaNGA sample,
yet nonetheless the LIRG and DYNAMO samples agree fairly well with our extrapolated
$\sigma_{\Ha} - \Sigma_{\rm SFR}$ relation
(though the 70 \kms\ ULIRG sample is still notably high).
In terms of total SFR the \citet{varidel16} sample (using their flux-weighted mean $\sigma_{\Ha}$ corrected for beam smearing)
agrees well with the extrapolated MaNGA relation, while the 
\citet{arribas14} LIRG/ULIRG data points are both substantially higher.

Finally, we note that an early analysis of the MaNGA data was presented by \citet{yu19}, who used 648 galaxies
from the MaNGA MPL-5 (DR14) data set and similarly noted 
positive correlations between velocity dispersion and SFR, $M_{\ast}$, and $\Sigma_{\rm SFR}$.
However, this study had significantly different methodology than our present contribution.
First, \citet{yu19} measured the \Ha\ velocity dispersion from stacked galaxy spectra, which included a substantial component
due to galactic rotation (although this was partially mitigated by their beam smearing correction and preferential selection of face-on spiral disks).  Second, as demonstrated in
\citet{paper1} the 
DR14 data products used by \citet{yu19} adopted an instrumental LSF estimate that was in error by about 5\%
in the vicinity of \Ha.  As a result, the 30-50 \kms\ values presented by \citet{yu19} were likely systematically
overestimated by about 15 \kms\ compared to the DR17 analysis of the full MaNGA
data set presented here.


\section{Secondary Relations}
\label{details.sec}

In \S \ref{results.sec} we confirmed the existence of a correlation between gas-phase velocity dispersion and both the local
star formation rate surface density and the galaxy-integrated total star formation rate, as expected on the basis of theoretical
work and recent observational studies.
However, there are multiple other correlations that can be physically expected as well;
for instance, since the most rapidly star forming galaxies on the main sequence tend to
be the most massive, we would expect to see a correlation between velocity dispersion
and stellar mass as well.  Likewise, artifacts produced by an imperfect beam smearing
correction would tend to be largest for the highest mass galaxies, potentially
masquerading as a correlation with the star formation rate.
In this section, we use the statistical power of MaNGA to investigate such correlations
and narrow down the most likely physical cause of the enhanced velocity dispersions.


\subsection{Redshift}
\label{redshift.sec}

The redshift range of the MaNGA sample ($z = 0.015 - 0.1$) is too small to expect any
significant physical evolution 
in the galaxy population, and therefore presents a good test of our ability to separate
genuine physical trends in $\langle \sigma_{\Ha} \rangle$ from trends imparted by
the correlation of $z$ with other observables.
Specifically, redshift is strongly correlated with both stellar mass and total SFR
\citep[see, e.g.,][their Figure 1]{paper2}
for the simple reason that massive, high-SFR disk
galaxies are rare and thus occur more frequently at larger redshifts corresponding
to a larger survey volume.  At the same time, our fixed angular resolution will correspond
to larger physical scales at higher redshifts, resulting in more significant beam smearing.

\begin{figure*}[!]
\epsscale{1.1}
\plotone{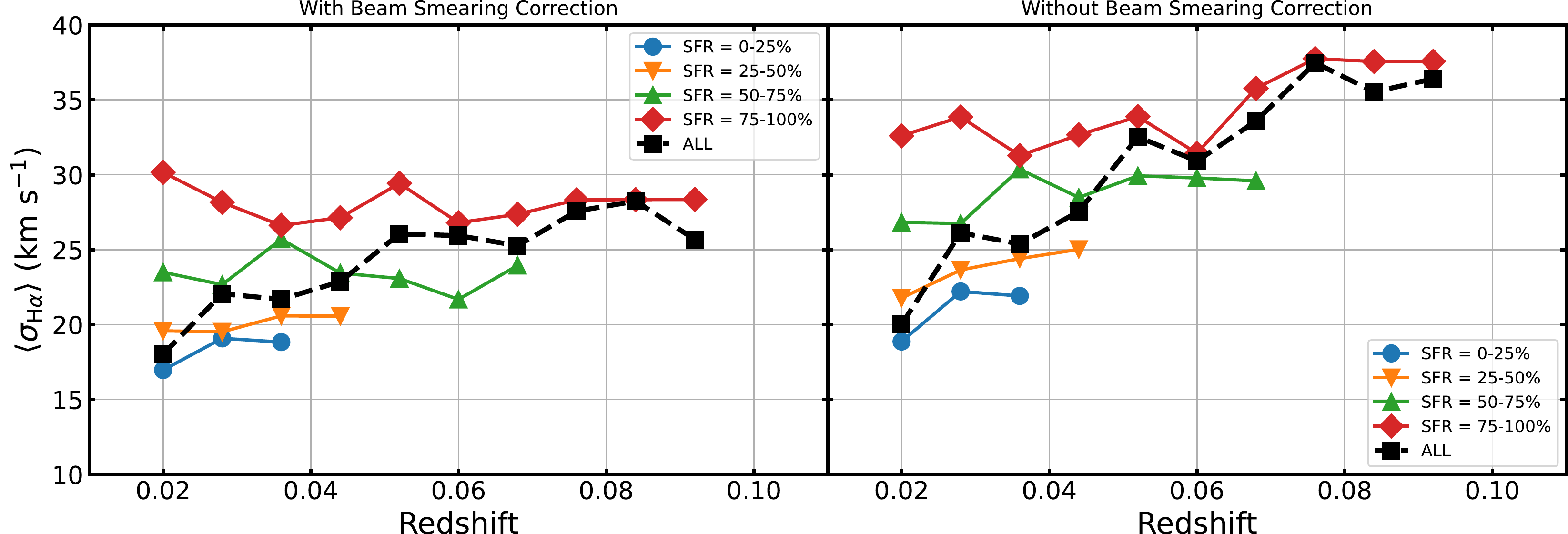}
\caption{Galaxy-averaged \Ha\ velocity dispersion with (left panel)
and without (right panel) beam smearing correction applied as a function of redshift for
all star-forming galaxies (dashed black line and points) and
for four bins in the total star formation rate (colored lines and points).
All points represent $2.5\sigma$-clipped means applied to the observational sample.
Statistical uncertainties in the mean for each point are $< 1$ \kms.
The apparent relation between $\langle \sigma_{\Ha} \rangle$ and redshift
after beam-smearing correction
is driven by the strong relation between redshift and total SFR.
}
\label{redshift.fig}
\end{figure*}

Both effects will tend to produce a relation in the sense that we expect 
$\langle \sigma_{\Ha} \rangle$ to increase as a function of redshift.  Indeed, this is
exactly what we see in Figure \ref{redshift.fig} (left panel, dashed black line).  However, this
apparent relation is driven by the correlation of $z$ with total SFR; if we
subdivide the sample into bins by SFR (Fig. \ref{redshift.fig}, left panel, colored lines) we note
that there is no relation between $\langle \sigma_{\Ha} \rangle$ and $z$ within a given
SFR bin.\footnote{In this and all later such figures, we plot values for a given bin only if there are a sufficient number of galaxies in that bin (typically $\gtrsim$ 50) that the statistics are reliable.  Nonetheless, low number statistics can still produce some single-point artifacts in these figures.}  In contrast, $\langle \sigma_{\Ha} \rangle$ increases consistently between
each bin in SFR, and the dominance of high-SFR galaxies at high-z produced the 
apparent relation.

Figure \ref{redshift.fig} also provides a sanity check on our beam smearing correction;
if there were a significant uncorrected effect from beam smearing we should expect
these relations to increase as a function of $z$ even within a given SFR bin, which is 
exactly what we see if we use values of $\langle \sigma_{\Ha} \rangle$ that have
not been corrected for beam smearing (right-hand panel).

\subsection{Stellar Mass}
\label{mass.sec}

\begin{figure*}[!]
\epsscale{1.0}
\plotone{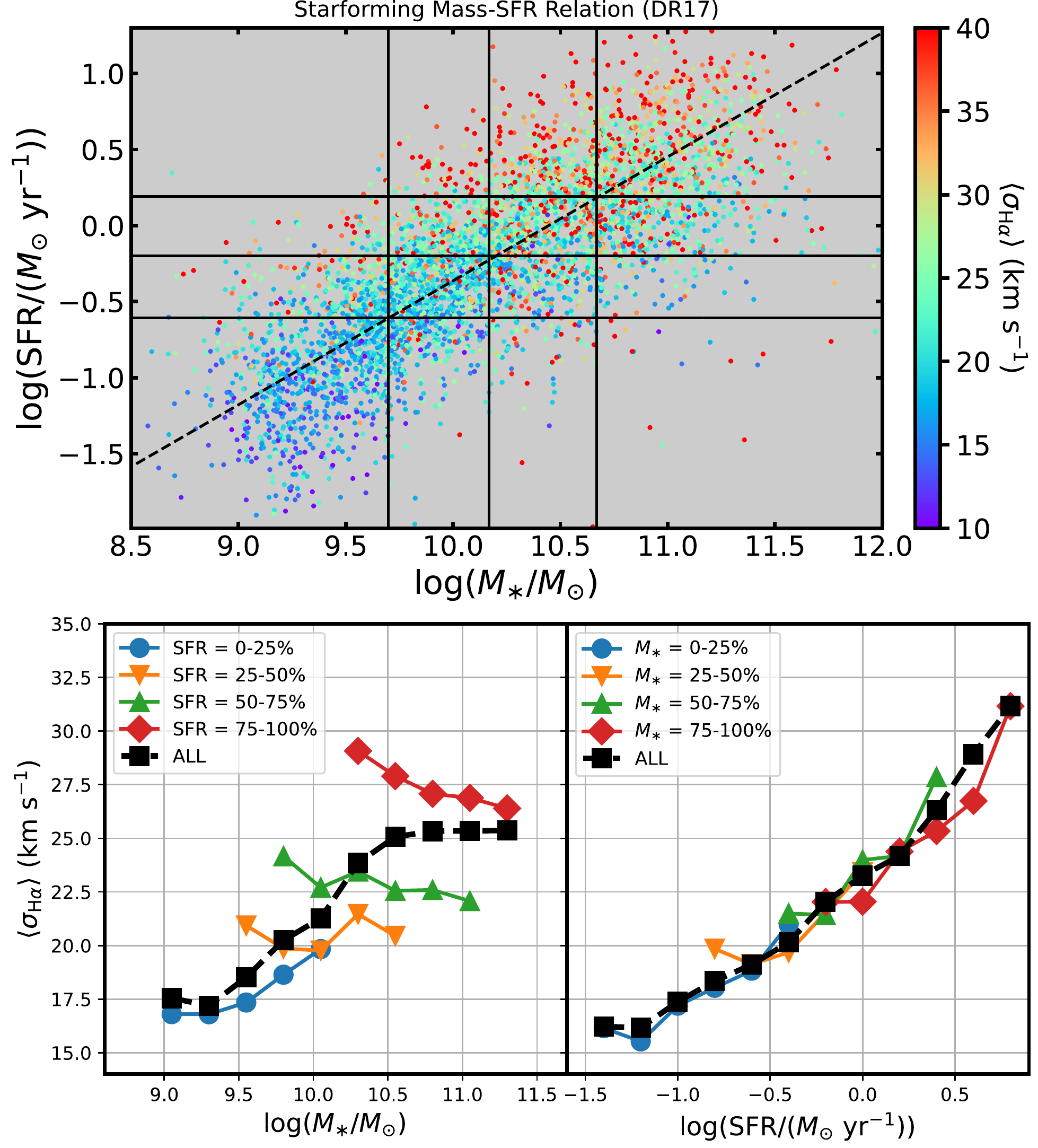}
\caption{Top panel: Total stellar mass \citep[derived from the NSA catalog,][]{blanton11} vs
total extinction-corrected \Ha\ star formation rate within the MaNGA footprint
for all 4517 galaxies in our star-forming sample, color-coded by their intensity-averaged
\Ha\ velocity dispersion.  Solid black lines show the corresponding quartile divisions
in both axes, the dashed black line shows a power-law fit to the MaNGA main sequence
using an orthogonal distance regression algorithm.
$\langle \sigma_{\Ha} \rangle$ is strongly correlated with location
along this star-forming main sequence.
Bottom-left panel: $\langle \sigma_{\Ha} \rangle$ as a function of stellar mass
for all points (dashed black line) and for four quartiles in SFR (colored lines).
Bottom-right panel: $\langle \sigma_{\Ha} \rangle$ as a function of SFR
for all points (dashed black line) and for four quartiles in stellar mass (colored lines).
All points represent $2.5\sigma$-clipped means applied to the observational data.
}
\label{masssfr.fig}
\end{figure*}

By definition, the star-forming galaxy main sequence describes a correlation between
stellar mass and star-formation rate, and the intensity averaged gas-phase velocity dispersion $\langle \sigma_{\Ha} \rangle$ is strongly correlated with the position
of a galaxy along this sequence.  As illustrated in Figure \ref{masssfr.fig} (top panel),
the lowest velocity dispersions $\langle \sigma_{\Ha} \rangle \sim 15$ \kms\ occur
at the lower end of this sequence, while the highest values $\sim 40$ \kms\ occur at the
upper end of the sequence.  However, to what extent is the increase due to the 
increasing stellar mass (increasing both the gravitational potential of the galactic disk
and the magnitude of shear within that disk)
vs the increasing star formation rate (increasing the amount of feedback injected into the
interstellar medium)?  Is it possible to distinguish which is the {\it primary} 
correlation, and which is simply a secondary correlation?

We endeavour to break this degeneracy by subdividing the galaxy sample according to
quartiles in both SFR and stellar mass, and plotting the residual correlations within
each quartile.  

In Figure \ref{masssfr.fig} (lower-left panel) we show the $2.5\sigma$-clipped mean
$\langle \sigma_{\Ha} \rangle$ as a function of stellar mass for the overall galaxy
sample (dashed black line), and for four quartiles in the total SFR (colored lines).
We note that the overall relation for all galaxies considered together
shows a strong positive correlation, and runs between 17 \kms\ and 25 \kms.
Within individual quartiles in total SFR however, this correlation with stellar
mass is almost entirely absent.
While the lowest-SFR bin (filled circles) increases slightly from 
17 - 20 \kms\ with increasing stellar mass, the second (downward-pointing triangles),
third (upward-pointing triangles), and fourth (diamonds) quartiles
in SFR show no such increase (and if anything mild evidence for a {\it decrease}) in
$\sigma_{\Ha}$ with increasing $M_{\ast}$.

If galaxies were homologous systems, which is generally a good first order approximation when studying galaxy scaling relations, and the $\langle \sigma_{\Ha} \rangle$ was tracing the gravitational potential, one would expect a $\langle \sigma_{\Ha} \rangle \propto \sqrt{M_{\ast}}$ dependence. This implies that if $\langle \sigma_{\Ha} \rangle = 25$ \kms\ at $10^{11} M_{\odot}$ one would expect 
$\langle \sigma_{\Ha} \rangle = 25/\sqrt{10} = 8$ \kms\ at
$10^{10} M_{\odot}$; even the overall global trend is clearly much more shallow than this.

In contrast, in Figure \ref{masssfr.fig} (lower-right panel) we show the
$2.5\sigma$-clipped mean
$\langle \sigma_{\Ha} \rangle$ as a function of SFR for the overall galaxy
sample (dashed black line), and for four quartiles in the total stellar mass (colored lines).
This overall relation is stronger than the $\langle \sigma_{\Ha} \rangle$ - $M_{\ast}$ relation,
running between 15 \kms\ and 30 \kms.  In addition, the correlation between
$\langle \sigma_{\Ha} \rangle$ and SFR very closely follows the average relation
in all four of the mass-quartile bins with no evidence of a systematic 
vertical offset between the bins.

We therefore conclude that the correlation between $\langle \sigma_{\Ha} \rangle$
and SFR is the {\it primary} relation, and that the apparent correlation with
stellar mass is driven by the preferential location of high-SFR objects at
higher stellar masses.

Numerous authors have similarly examined the relation between $\langle \sigma_{\Ha} \rangle$
and stellar mass in the past, with mixed results.
\citet{moiseev15} for instance concluded that the trend of $\langle \sigma_{\Ha} \rangle$
was stronger with \Ha\ luminosity than with stellar mass for their sample of dwarf galaxies,
as did \citet{varidel20} for the SAMI galaxy sample.
Unlike \citet{varidel20} however, we do not see evidence of a residual correlation
between $\langle \sigma_{\Ha} \rangle$ and $M_{\ast}$ after accounting for the SFR relation.
\citet{epinat10} did not observe a correlation between $\langle \sigma_{\Ha} \rangle$ and the maximum rotation velocity 
(i.e., a proxy for stellar mass) of galaxies in their sample, but noted that 
at higher redshifts (or lower spatial
resolution) beam smearing was capable
of artificially producing such a trend if uncorrected.  The importance of beam smearing
matches with our own observations; our $\langle \sigma_{\Ha} \rangle$ vs stellar
mass relation is nearly twice as strong if we were to instead use velocity dispersions
that have not been corrected for beam smearing, and less straightforward to disentangle
from the SFR relation.
Given how shallow the beam-smearing corrected trend with stellar mass is in the MaNGA data,
Monte Carlo tests drawing random subsamples of our data suggest that it is
unsurprising that \citet{epinat10} would have been unable to detect it in a sample
1/40th the size of MaNGA.


\subsection{Specific SFR, Main Sequence Offset, and Gas Fraction}
\label{gasfrac.sec}

\begin{figure*}[!]
\epsscale{1.2}
\plotone{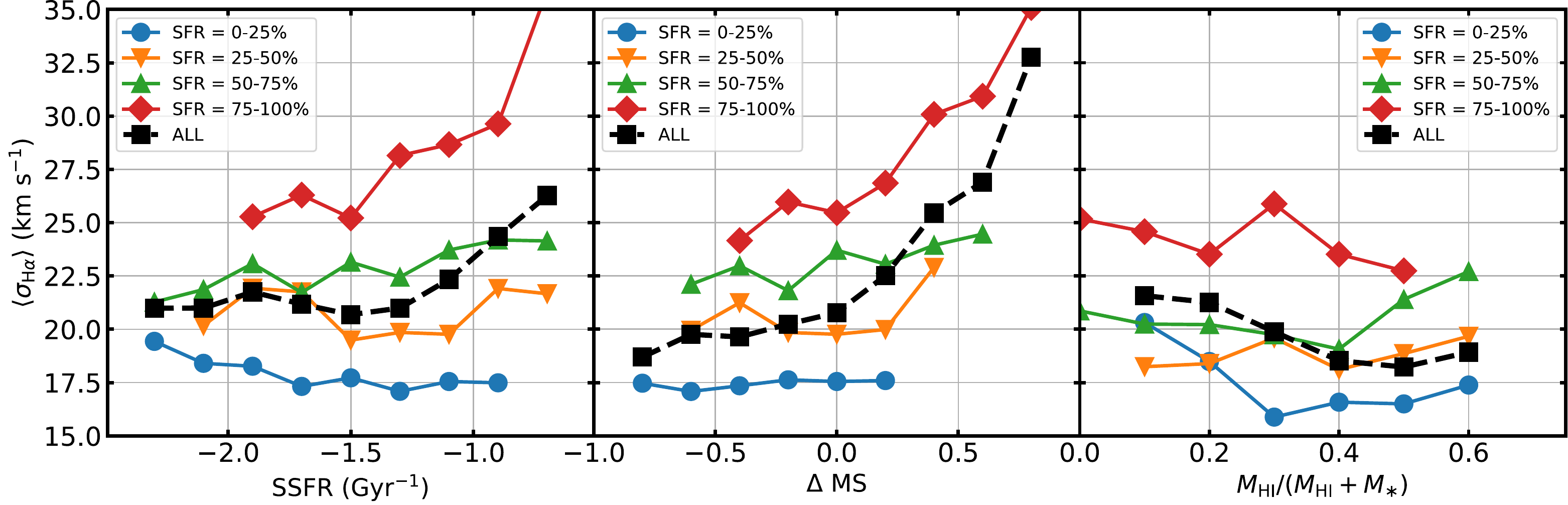}
\caption{Intensity-weighted mean \Ha\ velocity dispersion for galaxies in the
MaNGA sample as a function of their specific star formation rate (left-hand panel),
SFR offset from the main sequence (middle panel), and HI gas fraction
(right-hand panel).  HI gas fractions are based on 1732 galaxies of the sample
with HI gas masses drawn from \citet{stark21}.
The majority of the trends in all panels are best explained
by the vertical offset between individual quartiles of the galaxy sample
in total SFR (colored lines and points).
All points represent $2.5\sigma$-clipped means applied to the observational data.
}
\label{others.fig}
\end{figure*}

Similarly to \S \ref{mass.sec}, we can investigate whether any other variations on the SFR
are even more closely correlated to the ionized gas velocity dispersion.
Following \citet{varidel20}, we compute both the specific star formation rate
(SSFR, i.e. the total SFR divided by the stellar mass) and the main-sequence SFR offset
($\Delta$MS, i.e. the SFR `excess' of a given galaxy above the MaNGA-derived main sequence
mean for the stellar mass; see dashed line in Figure \ref{masssfr.fig}).
As indicated by Figure \ref{others.fig} we see correlations of $\langle \sigma_{\Ha} \rangle$ with both quantities.  However, these trends appear 
to again be be driven 
primarily by the underlying correlation with total SFR as suggested by the vertical
offset in the colored lines.  While there appears to be a residual trend within
the highest-SFR bin, this is an artifact of the wide range of this bin and
disappears if we subdivide further in total SFR (see, e.g., \S \ref{bigcor.sec}).

Additionally, we compute the HI gas fraction
using Green Bank Telescope and ALFALFA HI masses drawn from the ongoing HI-MaNGA survey \citep{masters19,stark21}.\footnote{Data are available as a Value Added Catalog in SDSS DR17.}
We find that 1732 of our galaxies have HI measurements flagged as reliable in
v2.0.1 of the HI-MaNGA catalog.  In Figure \ref{others.fig} (right-hand panel) we show
the intensity-weighted velocity dispersion as a function of the HI gas fraction,
and observe a small decline of $\sim 3-4$ \kms\ from 0-60\% gas fraction that is statistically
significant at $8\sigma$ based on the error in the mean.  \citet{varidel20}
noted a similar relation in their SAMI data, although these authors were unable to determine
if the relation was significant.
This relation, however, appears to be a
consequence of the anticorrelation between gas fraction and total SFR in the sense
that the largest gas fractions occur in the lowest mass galaxies with low total SFR.
As indicated by the colored lines in Figure \ref{others.fig} (right-hand panel), no such trends
are convincing within individual SFR bins.


\subsection{SFR Surface Density}
\label{bigcor.sec}

Finally, we investigate whether we can break the correlation
between local star formation rate surface density $\Sigma_{\rm SFR}$ and the total
galaxy-integrated SFR in order to determine whether local or global properties
are the physical driver of the observed relation.
The underlying tension between local and global galaxy properties and their respective influence on
star formation has been a subject of substantial debate for many years
\citep[see, e.g.,][and references therein for a recent review]{sanchez21a,sanchez21b}.

\begin{figure*}[!]
\epsscale{1.1}
\plotone{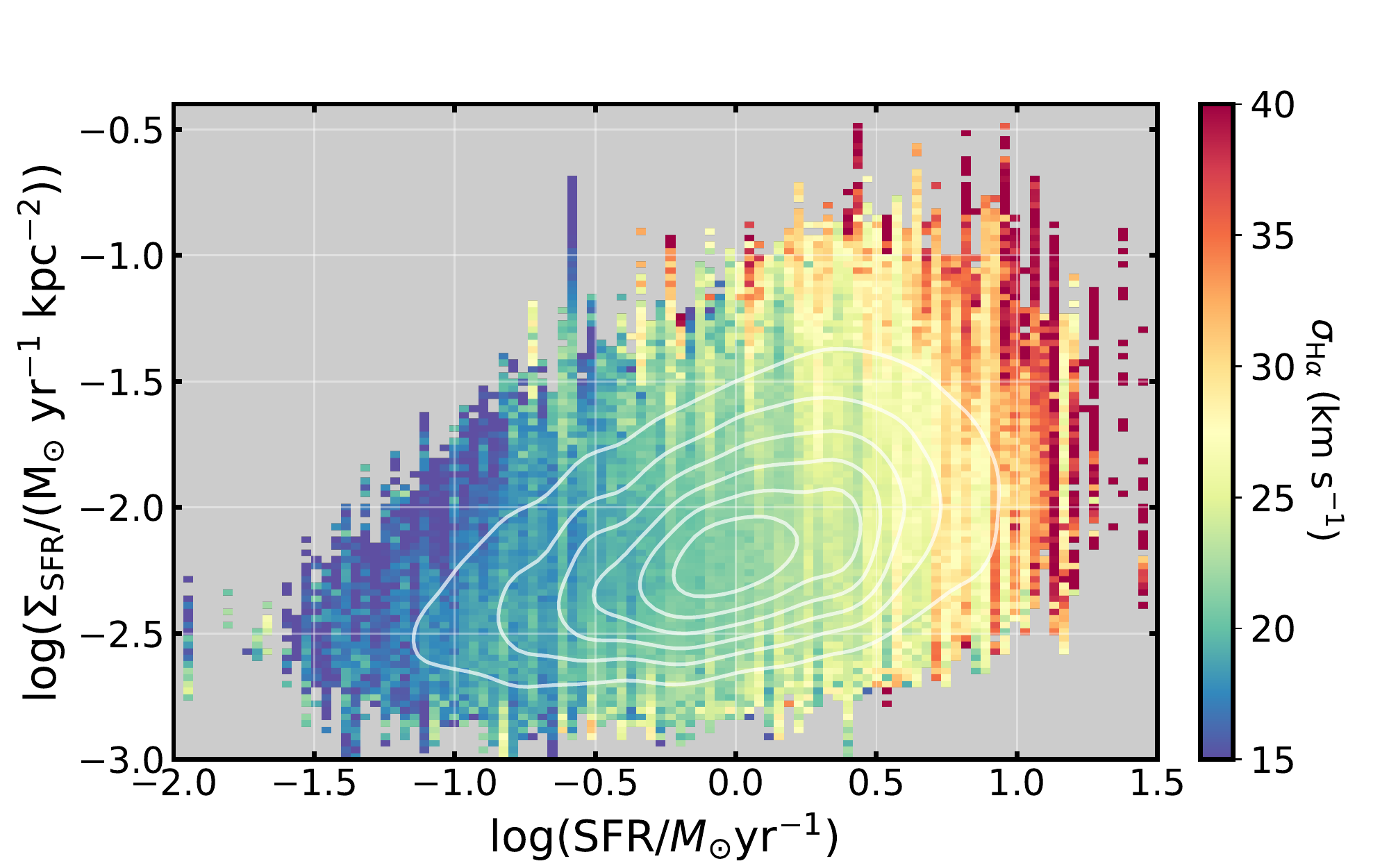}
\caption{Local star-formation rate surface density $\Sigma_{\rm SFR}$ for individual
MaNGA spaxels vs the total SFR of the galaxy within which they reside, color-coded
by the $2.5\sigma$-clipped mean \Ha\ velocity dispersion of the spaxels at each point.
Solid white contours indicate the linear number density distribution of spaxels
within the plot.
Binning artifacts in the SFR axis are caused by the discrete nature of the integrated
SFR for the 4517 galaxies in the MaNGA sample compared to the 1.4 million individual
spaxel measurements.  High-$\Sigma_{\rm SFR}$ spaxels are located almost exclusively
in high-SFR galaxies.
}
\label{sfr_sigmasfr1.fig}
\end{figure*}

Unsurprisingly, as indicated by Figure \ref{sfr_sigmasfr1.fig}
$\Sigma_{\rm SFR}$ and total SFR are closely correlated with each other in the sense that the
regions of highest SFR surface density ($\Sigma_{\rm SFR} \sim 10^{-1.5} M_{\odot}$ yr$^{-1}$ kpc$^{-2}$) tend to live within the most rapidly star-forming
galaxies.  There is nonetheless a wide range though, with
regions of $\Sigma_{\rm SFR} = 10^{-2.5} M_{\odot}$ yr$^{-1}$ kpc$^{-2}$
occurring in galaxies
of all SFR.  Thus, it should be possible to test whether the velocity dispersion is higher near
regions of active star formation in lower-SFR galaxies or far away from regions of active 
star formation in higher-SFR galaxies.

\begin{figure*}[!]
\epsscale{1.1}
\plotone{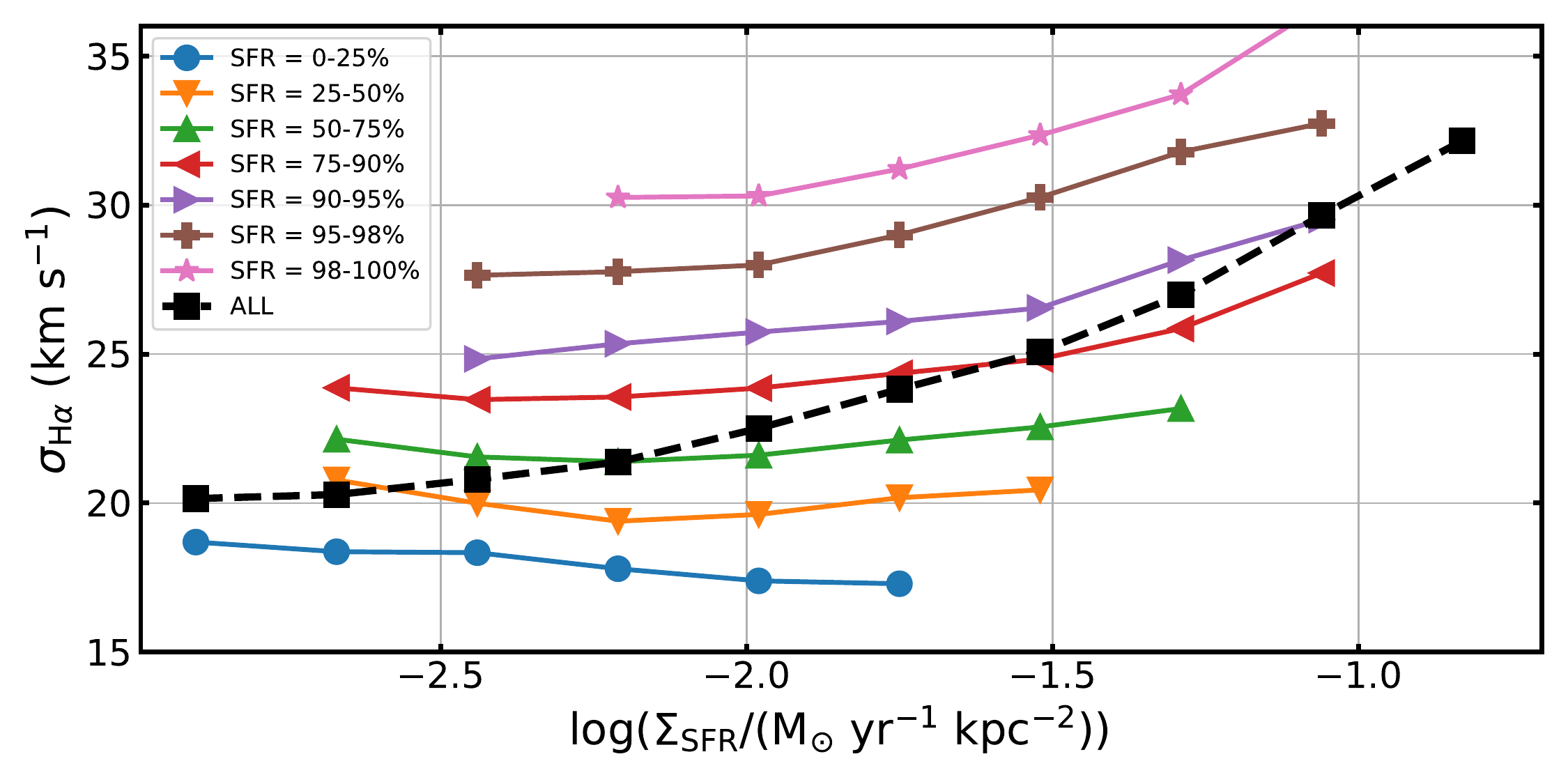}
\caption{Spaxel velocity dispersion as a function of the local star formation rate
surface density $\Sigma_{\rm SFR}$ for a variety of bins in total galaxy SFR (colored
points and lines).  The mean relation for all galaxies (dashed black line; see
Figure \ref{sigma_sfrsd.fig}) is produced primarily by the offset between individual
SFR subsamples, which are nearly flat as a function of $\Sigma_{\rm SFR}$.
Individual points represent $2.5\sigma$-clipped averages of the observational data.
}
\label{sfr_sigmasfr2.fig}
\end{figure*}

Figure \ref{sfr_sigmasfr1.fig} demonstrates the results relatively convincingly on its own; the sigma-clipped
mean $\sigma_{\Ha}$ as a function of location within the diagram increases strongly with SFR, but less convincingly with $\Sigma_{\rm SFR}$.
We dissect this furter in Figure \ref{sfr_sigmasfr2.fig}, breaking the relation between 
$\Sigma_{\rm SFR}$ and $\sigma_{\Ha}$ into multiple bins of total SFR, with 
Table \ref{percentiles.table} giving
the translation of percentile ranges from the MaNGA sample to actual SFR.
We observe that this relation appears to be governed primarily by the correlation
between $\Sigma_{\rm SFR}$ and total SFR; at fixed $\Sigma_{\rm SFR}$ there is a strong
increase in $\sigma_{\Ha}$ as a function of total SFR (from 17-30 \kms\ at 
$\Sigma_{\rm SFR} = 10^{-2} M_{\odot}$ yr$^{-1}$ kpc$^{-2}$).  In contrast, $\sigma_{\Ha}$
is nearly constant over two orders of magnitude as a function of $\Sigma_{\rm SFR}$ at fixed SFR.

Although
there may be a small positive relation with $\Sigma_{\rm SFR}$ in higher-SFR bins (visible also
as slight vertical gradients in $\sigma_{\Ha}$ in Figure \ref{sfr_sigmasfr1.fig}), the majority
of such trends are produced by
residual correlations between $\Sigma_{\rm SFR}$ and SFR {\it within} each SFR bin.
If, for instance, we had plotted a single bin of galaxies with SFR between the 75th and 100th
percentiles we would have observed a strong correlation between $\Sigma_{\rm SFR}$
and $\sigma_{\Ha}$.  Splitting this out as we have done into 75th-90th percentile, 90th-95th
percentile, etc. we see that the trend is driven more by the total SFR.

We therefore conclude that the increase of ionized gas velocity dispersion is driven
more by the global SFR of a galactic disk than by local
properties governing the injection of supernova feedback at the sites of peak star formation
within that disk.  We caution, however, that the SFR itself is likely not the underlying physical
driver of the observed relation, and some other as-yet untested global quantity that correlates strongly with total SFR (e.g., 
the total molecular gas mass) may be the true cause.

One potential explanation for this may be that (at least at kpc scales) velocity dispersions are set largely by the global increase of the disk midplane pressure in the presence of larger quantities of cold gas, and correspondingly higher star formation rates.
Indeed, both \citet{hughes13} and \citet{sun20} noted that giant molecular clouds appear to `know'
about the global properties of their host galaxies, with galaxies with higher stellar masses and SFR
hosting molecular gas with higher surface densities and velocity dispersions.
As we demonstrate in \S \ref{molec.sec}, after accounting for physics internal to HII regions the velocity dispersion of the molecular gas disk in which HII regions
are embedded increases with total molecular gas mass density in agreement with
recent ALMA CO (2-1) observations.
While localized phenomena such as spiral arms and bars tend to be associated with regions of enhanced
velocity dispersion as well both in molecular gas observations \citep[e.g.,][]{sun20} and in theoretical 
models \citep[e.g.,][]{nguyen18}, these effects may simply be difficult to discern at the dynamic range and kpc-scale resolutions to which MaNGA is sensitive (although see discussion in \S \ref{inclin.sec}).


Our results therefore take a step towards understanding
why some studies have found a relation between 
$\Sigma_{\rm SFR}$ and $\sigma_{\Ha}$ at higher redshifts \citep[e.g.,][]{swinbank12,lehnert13} 
while others have not \citep[e.g.,][their Fig 8]{ubler19}.  With low number statistics and coarse spatial resolution, the strength
of this relation will be driven in large part by the range 
and distribution of SFR in the galaxy sample instead of
the dynamic range of the observations within any individual galaxy.
Likewise, the findings from \citet{colina05} and \citet{arribas14} that
localized peaks in velocity dispersion don't tend to be correlated
with peaks in $\Sigma_{\rm SFR}$ in local LIRGs/ULIRGs
are naturally explained if $\Sigma_{\rm SFR}$ is not the primary driver of enhanced velocity dispersions.

\begin{deluxetable}{cc}
\label{percentiles.table}
\tablecolumns{2}
\tablewidth{0pc}
\tabletypesize{\scriptsize}
\tablecaption{SFR Percentiles}
\tablehead{
\colhead{Percentile} & \colhead{log(SFR/$M_{\odot}$ yr$^{-1}$)}
}
\startdata
0\% & -2.02\\
25\% & -0.61\\
50\% & -0.20\\
75\% & 0.19\\
90\% & 0.51\\
95\% & 0.70\\
98\% & 0.90\\
100\%  & 2.09
\enddata
\end{deluxetable}

\section{Disk Inclination and Azimuthal Angle}
\label{inclin.sec}

Thus far, we have made no explicit cuts to the galaxy sample in terms of the
inclination $i$ of the galactic disk to the line of sight,
nor any corrections for
inclination-dependent effects.  Indeed, the full MaNGA galaxy sample  was
selected in a manner agnostic to inclination, although there is a small bias
towards more face-on galaxies ($i =0^{\circ}$)
at a given stellar mass because
additional extinction in edge-on disks ($i =90^{\circ}$) tends to give fainter
optical magnitudes corresponding to a smaller redshift volume \citep[see discussion by][]{wake17}.  As such, MaNGA galaxies span a wide range of inclinations
from $i = 0^{\circ} - 90^{\circ}$ with mean and median values of 
56$^{\circ}$ and 57$^{\circ}$ respectively, 
consistent with expectations for a population of randomly-oriented
disks 
\citep[$57.3^{\circ}$ and $60^{\circ}$ respectively; see derivation in Appendix A of][]{law09}.

Here, we have estimated the inclination of our star-forming disk sample based on the $r$-band Sersic profile
morphological minor and major axis lengths given by the NSA catalog \citep{blanton11}
from which the MaNGA parent catalog was derived.  Assuming an intrinsic axis ratio of $q_0 = 0.15$
\citep[e.g.][]{ryden06},
the inclination is given by \citep{holmberg58}:
\begin{equation}
    \textrm{cos}^2 i = \frac{(b/a)^2 - q_0^2}{1-q_0^2}
\end{equation}
where $a$ and $b$ are the major and minor axis lengths respectively.\footnote{In practice,
assuming $q_0 = 0.15$ instead of $q_0 = 0$ only makes a difference of $> 2^{\circ}$ in the
recovered inclination for $i > 70^{\circ}$.}

In the past, many studies have opted to focus on face-on disk galaxies in order to 
ensure that the the observed line-of-sight velocity dispersion $\sigma_{\rm los}$
is nearly the disk vertical velocity dispersion ($\sigma_{\rm z}$) and minimize confusion from
the radial ($\sigma_{\rm r}$) and azimuthal ($\sigma_{\phi}$) components, 
along with minimizing observational biases from beam smearing that are more severe
at large inclinations \citep[see, e.g.,][]{cappellari20}.
Using the statistical power of the MaNGA sample, we quantify the magnitude of such effects
at kpc-scale resolution for the star-forming main sequence galaxy distribution at $z=0$.

\begin{figure*}[!]
\epsscale{1.2}
\plotone{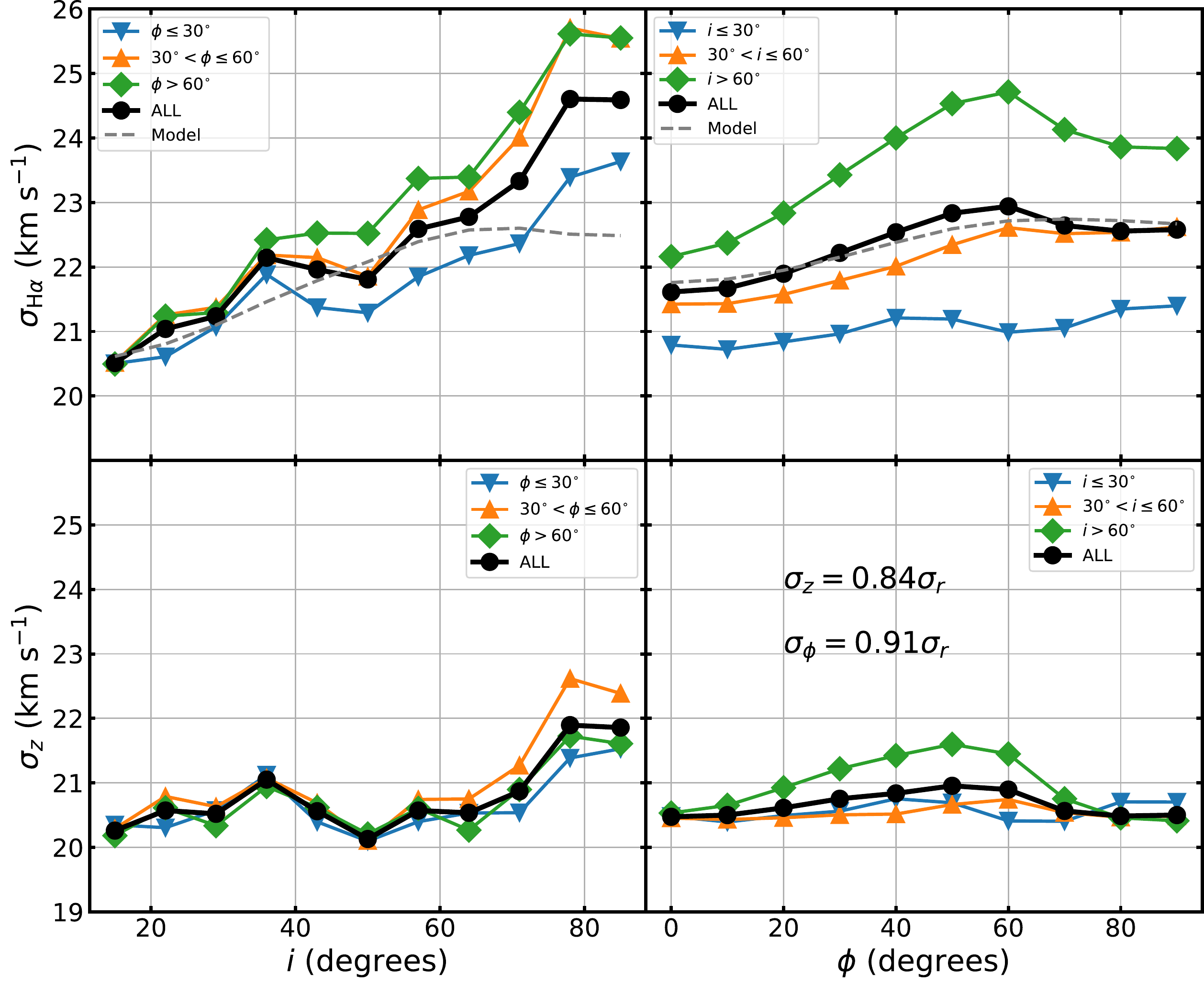}
\caption{Top panels: Line-of-sight velocity dispersion $\sigma_{\rm los}$ (defined equal to the
observed $\sigma_{\Ha}$) as a function of galaxy inclination $i$ and spaxel
azimuthal angle $\phi$ for various ranges of both (solid black and colored lines respectively).   
$i = 0^{\circ}/90^{\circ}$ corresponds to face-on/edge-on systems respectively, while $\phi = 0^{\circ}/90^{\circ}$ corresponds to locations on the major/minor axis respectively.
Dashed grey lines
indicate the expected value of $\sigma_{\Ha}$ given by Equation \ref{sigz.eqn} with the best-fit choices for $\sigma_z/\sigma_r$ and $\sigma_{\phi}/\sigma_r$; the turnover in this relation in the top-left panel is due to the preferential bias of MaNGA spaxels towards the major axis for the most edge-on galaxies.
Bottom panels: Disk vertical
velocity dispersion $\sigma_z$ as a function of galaxy inclination $i$ and spaxel
azimuthal angle $\phi$ for an appropriate choice of $\sigma_z = 0.85 \sigma_r$ and $\sigma_{\phi} = 0.91 \sigma_r$.
All points represent $2.5\sigma$-clipped means applied to the observational data.
}
\label{inclinfig.fig}
\end{figure*}

In Figure \ref{inclinfig.fig} (top panels) we plot our observed $\sigma_{\Ha} \equiv \sigma_{\rm los}$ as a function of both the galaxy inclination angle $i$ and the disk
azimuth angle $\phi$ of each spaxel (where the values reported by the DAP have been collapsed via symmetry to the range $\phi = 0^{\circ}-90^{\circ}$, such that $\phi = 0^{\circ}$
and $\phi = 90^{\circ}$ correspond to the disk major and minor axes respectively, as
given by the NSA catalog).
In each plot, we further subdivide the points according to bins in the other parameters;
i.e., we plot $\sigma_{\Ha}$ as a function of $i$ for various bins in $\phi$,
and as a function of $\phi$ for various bins in $i$.

We note that the overall $\sigma_{\Ha}$ increases by about 3 \kms\ 
as a function of $i$
for all $\phi$, albeit slightly offset from each other.\footnote{Simplistically, the observed
line of sight velocity dispersion increases
by about 1 \kms\ for every 30$^{\circ}$ of inclination, suggesting a correction
factor of 1.5 \kms\ at the mean inclination of our sample.}  Likewise, $\sigma_{\Ha}$
increases by about 1 \kms\ as a function of $\phi$ overall, with both a difference in slope
and a vertical offset for different ranges in $i$.

\begin{figure}[!]
\epsscale{1.2}
\plotone{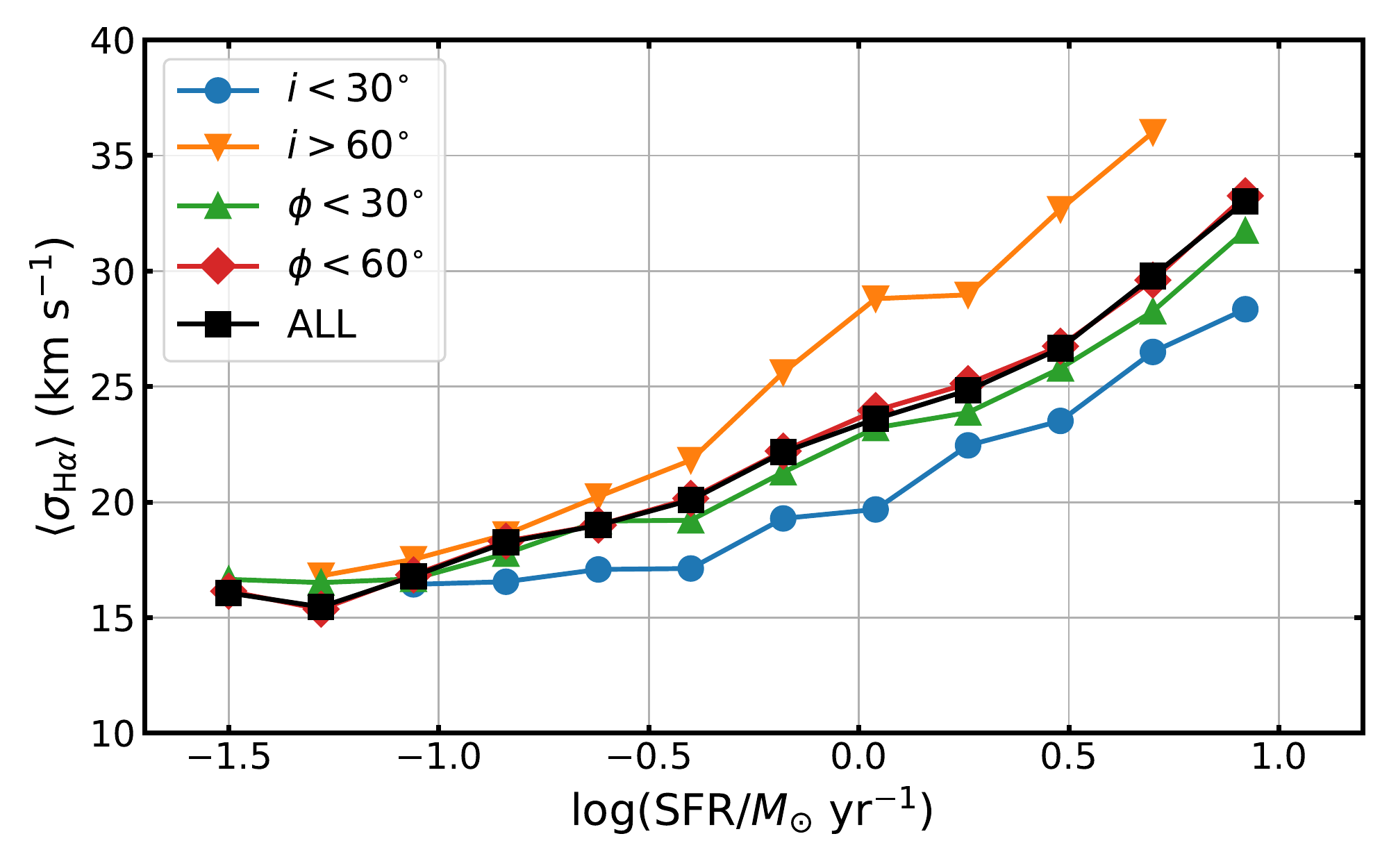}
\caption{Intensity-weighted mean velocity dispersion as a function of total dust-corrected
star formation rate for the entire galaxy sample (black line, c.f. Figure \ref{paper3_global.fig})
and for subsamples in galaxy inclination $i$ and spaxel azimuthal angle $\phi$
(colored lines).  Although inclination of the galaxy sample shifts the overall trend slightly,
it is nonetheless strong in all subsamples.
All points represent $2.5\sigma$-clipped means applied to the observational data.
}
\label{inclin2.fig}
\end{figure}

At the most simplistic level, we assess the potential bias that these trends have on our
derived relation between $\langle \sigma_{\Ha} \rangle$ and SFR by plotting the relation for
sub-populations of inclination and azimuthal angle in Figure \ref{inclin2.fig}.
As expected from the relative amplitude of the trends with $i$ and $\phi$, such effects
represent only a minor perturbation to the overall relation.

However, we can also go a step further by noting that changes in 
$\sigma_{\Ha}$
are produced by the inclusion of radial ($\sigma_r$) and azimuthal ($\sigma_{\phi}$)
velocity dispersions into the line
of sight projected velocity dispersion that was dominated by the disk vertical velocity
dispersion $\sigma_{\rm z}$ for face-on galaxies.
Both radial and azimuthal components increase in importance as $i$ increases; for larger inclinations
$\sigma_{\phi}$ becomes dominant on the major axis ($\phi < 30^{\circ}$) while
$\sigma_r$ becomes dominant on the minor axis ($\phi > 60^{\circ}$).

Mathematically, for a cylindrical alignment of the velocity ellipsoid (which is appropriate in the disk plane) the combination of these effects is given by geometric projection
\citep[see, e.g.][their Eqn. 27]{cappellari20} as\footnote{Note that this formalism breaks down at very large inclinations.}:
\begin{equation}
\sigma_{\rm los}^2 = \left( \sigma_r^2 {\rm sin}^2\phi + \sigma_{\phi}^2 {\rm cos}^2 \phi \right) {\rm sin}^2 i + \sigma_z^2 {\rm cos}^2 i
\label{jeans.eqn}
\end{equation}

Following \citep[][their Eqn. 10]{martinsson13}, we define $\alpha = \sigma_z/\sigma_r$ and $\beta = \sigma_{\phi}/\sigma_r$, in which case Equation \ref{jeans.eqn} can be rearranged as
\begin{equation}
    \sigma_z^2 = \frac{\sigma_{\rm los}^2}{{\rm cos}^2 i} \left[1 + \frac{{\rm tan}^2  i }{\alpha^2} ({\rm sin}^2\phi + \beta^2 {\rm cos}^2\phi) \right]^{-1}
\label{sigz.eqn}
\end{equation}

Making the approximate assumption of axisymmetry, which is appropriate for spiral galaxies,
the intrinsic components of the velocity ellipsoid
$\sigma_z$, $\sigma_r$, $\sigma_{\phi}$ (or equivalently, $\sigma_z$, $\alpha$, and $\beta$)
must be constant at a given galactocentric radius in the disk plane. If one further assumes that the gas kinematics of the galaxy population shares the same anisotropy $\alpha$ and $\beta$, one can use the variations of the projected $\sigma_{\rm los}$ at different inclinations and azimuthal angles to measure $\alpha$ and $\beta$ from the data. Here we determine these parameters by trying to remove any dependency of $\sigma_z$ on both inclination and azimuth angle.

\begin{figure}[!]
\epsscale{1.2}
\plotone{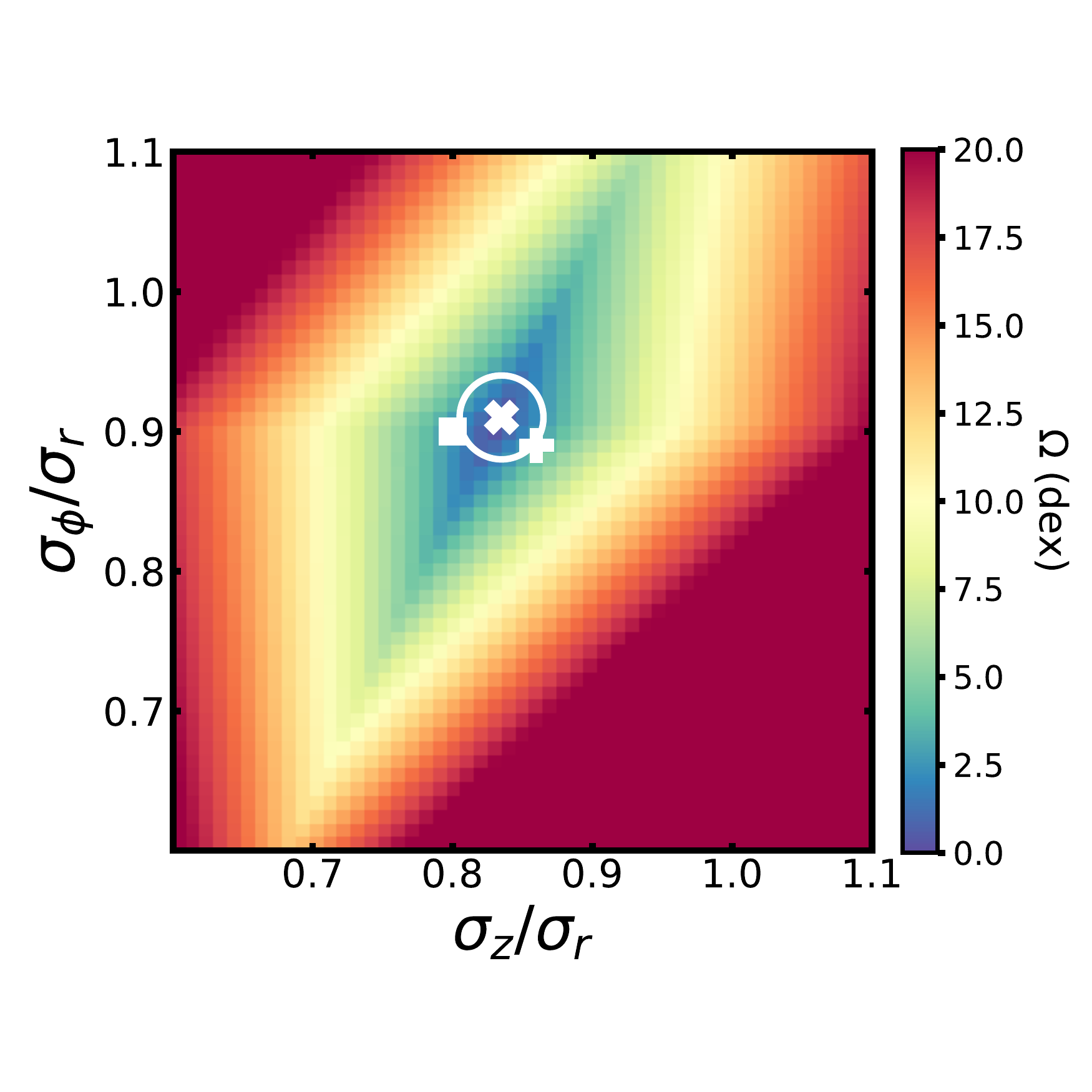}
\caption{Quality of fit statistic $\Omega$ describing the slope of the relation
between $\sigma_z$, $i$, and $\phi$ for a various choices of the velocity dispersion
ratios
$\sigma_z/\sigma_r$ and $\sigma_{\phi}/\sigma_r$.  The global minimum is prominently defined
in the region indicated by the white X, with the white circle indicating the $1\sigma$ uncertainty
in the measurement.  The white $+$ and white square indicate the best-fit values for the
MaNGA Primary and Secondary samples respectively.
}
\label{inclin_color.fig}
\end{figure}

We approach this problem numerically by considering
a grid of $\alpha$, $\beta$ in the range $0.6 - 1.1$ sampled every $0.01$ dex; for each grid point
we compute the slope of $\sigma_z$ as a linear function of $i$ and of $\phi$,
and sum the absolute values of these two slopes to compute a `flatness' statistic $\Omega$, the uncertainty of which
is given by the quadratic sum of the uncertainties in each of the two slopes.
In Figure \ref{inclin_color.fig} we show this surface in $\Omega$ and note that there is a well-defined
global minimum at 
\begin{equation}
    \sigma_z/\sigma_r = 0.84 \pm {0.03}
\end{equation} and
\begin{equation}
    \sigma_{\phi}/\sigma_r = 0.91 \pm 0.03
\end{equation}
Using these parameters, we observe that $\sigma_z$ is indeed flat as a function
of both $\phi$ and $i$ for all subsamples in which $i < 75^{\circ}$ (Figure \ref{inclinfig.fig}, lower panels).  At $i \geq 75^{\circ}$ our thin disk assumptions break down, and projection effects 
due to finite disk thickness and substructure
along a given line of sight become more significant, resulting in marginally higher values for $\sigma_z$.

This result is dependent on whether there is any residual beam smearing in our spaxel
sample; if we had not applied a beam smearing correction we would have derived nearly twice
the slope for $\sigma_{\Ha}$ as a function of $i$ or $\phi$.  The absence of any trend
between $\langle \sigma_{\Ha} \rangle$ and redshift (see \S \ref{redshift.sec}) suggests
that any such residual effect should be small; however, we confirm this by repeating our exercise using the
MaNGA Primary and Secondary galaxy samples which were designed to reach 1.5 and 2.5 effective
radii respectively and have median redshifts of $z = 0.027$ and $z = 0.05$
\citep[see Figure 1 of][]{paper2}.
While the Secondary sample has velocity dispersions that are about 1.5 \kms\ higher
than the Primary sample on average (likely due to the greater prevalence of high-SFR
objects in the larger cosmological volume), the derived
values of $\alpha = 0.86/0.80$ and $\beta = 0.89/0.90$ for the Primary/Secondary samples
respectively.  All four values are within about $1\sigma$ uncertainty of our estimate
from the whole MaNGA sample, although the subsample
$\alpha$ values differ from each other by $2\sigma$.  This difference may suggest some residual beam smearing, but may also
simply reflect the statistical uncertainty in our measurement.

Our result for the average velocity dispersion ellipsoid of the ionized gas disk in MaNGA galaxies is 
broadly in keeping with other estimates in the literature.
Most immediately, \citet{varidel20} estimated $\sigma_z/\sigma_r = 0.80^{0.06}_{-0.05}$ 
on the basis of 
383 galaxies from the SAMI survey.  Although these authors fixed $\sigma_{\phi}/\sigma_r = 1.0$,
their result for $\sigma_z/\sigma_r$ is consistent with ours to within $1\sigma$.
Similarly,
\citet[][see their Fig. 21]{leroy08} found that $\sigma_{\rm los}$ for
HI gas in the THINGS survey increased
by about 2 \kms\ from $i = 0 - 60^{\circ}$ (albeit within their quoted error
bars) with a significant spike at $i > 60^{\circ}$ that they ascribe to projection effects, consistent with our Figure \ref{inclinfig.fig}.

Likewise, \citet{guiglion15} estimate $\sigma_r$, $\sigma_{\phi}$, and $\sigma_z$ for
stellar populations in the Milky Way's Galactic disk using GAIA-ESO spectroscopic observations.  Their Figure 12 suggests that $\sigma_z/\sigma_r = 0.61$ and $\sigma_{\phi}/\sigma_r = 0.71$,
and $\sigma_z = 20$ \kms\ for the metal-rich (i.e., young) thin stellar disk.
A more recent analysis by \citet{nitschal21} using a combination of GAIA EDR3 and
SDSS-IV APOGEE spectroscopic observations found similar results, with 
 $\sigma_z/\sigma_r = 0.663$ and $\sigma_{\phi}/\sigma_r = 0.711$ (see their Table 1).

Although we should not
expect the gas disk and the young stellar disk to have identical kinematics,
it is nonetheless
suggestive that we find similar values for the vertical velocity dispersion, that
the radial velocity dispersion is significantly larger than the vertical dispersion, and that
the azimuthal dispersion is intermediate between the two.
Indeed, since young stars must have formed from HII regions relatively recently and both
broadly trace galactic features such
as bars and spiral arms, it is perhaps unsurprising that the trends are in general agreement.

We tested this further using morphological classifications drawn from the Galaxy Zoo
project \citep{willett13}, in which we describe the strength of the bar and/or spiral arms by the level
of agreement between individual classifiers on the presence or absence of such
a feature \citep[see, e.g.,][]{geron21}.  Our derived values for the ionized gas velocity dispersion ellipsoid in
strong vs weak or absent bars/arms\footnote{Defined as the top/bottom quartiles respectively for agreement on
the existence of such features.}
show inconclusive deviations from the main galaxy sample, although
our value of $\beta = \sigma_{\phi}/\sigma_r = 1.0$ for the strong-bar subsample
is marginally significant at $3\sigma$.  The strong-bar subsample also has 
line-of-sight velocity dispersions
that are larger than the unbarred subsample by about 2 \kms\ in the most highly-inclined
galaxies, suggesting that bars increase the azimuthal velocity dispersion
between HII regions on kpc scales.  While
galaxies with a strong spiral pattern also have velocity dispersions
that are 1-2 \kms\ larger than those without a clear spiral pattern, this is likely
driven by the increased prevalence of strong spiral patterns
in higher-mass galaxies \citep[a trend originally established by][]{elmegreen87}.
We investigate both trends in greater detail in a forthcoming contribution.


\section{Discussion}
\label{discussion.sec}


\subsection{Internal dynamics of the HII regions}
\label{hiiphysics.sec}

Far from being unbiased tracers of a uniform `ionized gas' layer, HII regions are distributed
according to wherever the local molecular gas density is high enough that it has been
able to condense and initiate star formation.  This star formation in turn has dramatic
effects on the surrounding gas, with the flood of ionizing photons
both heating the
gas and combining with mechanical feedback from the 
bright O+B stars to stir additional turbulence in the now-bright HII region
\citep[see, e.g., discussion by][]{osterbrock06}.
The velocity dispersions $\sigma_{\Ha}$ that we have measured from observations of the bright \Ha\ emission line will thus be a combination of thermal pressure within individual
HII regions, turbulent motions within those regions, and an overall dispersion between
individual HII regions that tells us about the distribution of molecular clouds within
the galactic disk.

The thermal pressure in such regions will be determined by their temperatures
$T \sim 10^4$K, corresponding to a thermal velocity dispersion
$\sigma_{\rm Therm} \approx 10$ \kms\ \citep[see, e.g.,][]{krumholz16}.
The non-thermal component of the velocity dispersion
$\sigma_{\rm Turb}$ is
broadly related to the expansion of the HII region, although the velocity profiles
in nearby galaxies may be more accurately described as turbulent
\citep{osterbrock06}.  The strength of this turbulent component may vary
significantly between different HII regions and contribute to the observed relation between
$\sigma_{\Ha}$ and total SFR.
Indeed, for giant HII regions there is a well-known relation $L_{\Hb} \propto \sigma_{\Hb}^5$ between
the turbulent velocity dispersion and the
total \Hb\ luminosity of the nebula 
that is sufficiently well-calibrated
\citep[see, e.g.,][]{larson81,melnick21}
as to be used for cosmological distance measurements.\footnote{Although note, however, that
some recent studies ascribe this relation in part to an observational artifact
resulting from distance-dependent resolution bias in GMC deconvolution algorithms
\citep{hughes13}.}
For typical HII regions however (many of which we expect to contribute to the signal in a given kpc-scale MaNGA beam)
values of $\sigma_{\rm Turb}^2$ range from 6-16 \kms\ across 4 orders of magnitude
in \Ha\ luminosity \citep{zc15}.  Since we observed no strong trends in $\sigma_{\Ha}$ with
$\Sigma_{\rm SFR}$ at fixed total SFR (see Figures \ref{sfr_sigmasfr1.fig} and \ref{sfr_sigmasfr2.fig}), we therefore follow 
\citet{osterbrock06} and \citet{krumholz16}
in assuming that 
$\sigma_{\rm Turb} \approx 10$ \kms\ on average.
Altogether, we expect that processes internal to the HII regions
themselves thus contribute $\sqrt{\sigma_{\rm Therm}^2 + \sigma_{\rm Turb}^2} = 14$
\kms\ to our \Ha\ velocity dispersion measurements.

\begin{figure*}[!]
\epsscale{1.2}
\plotone{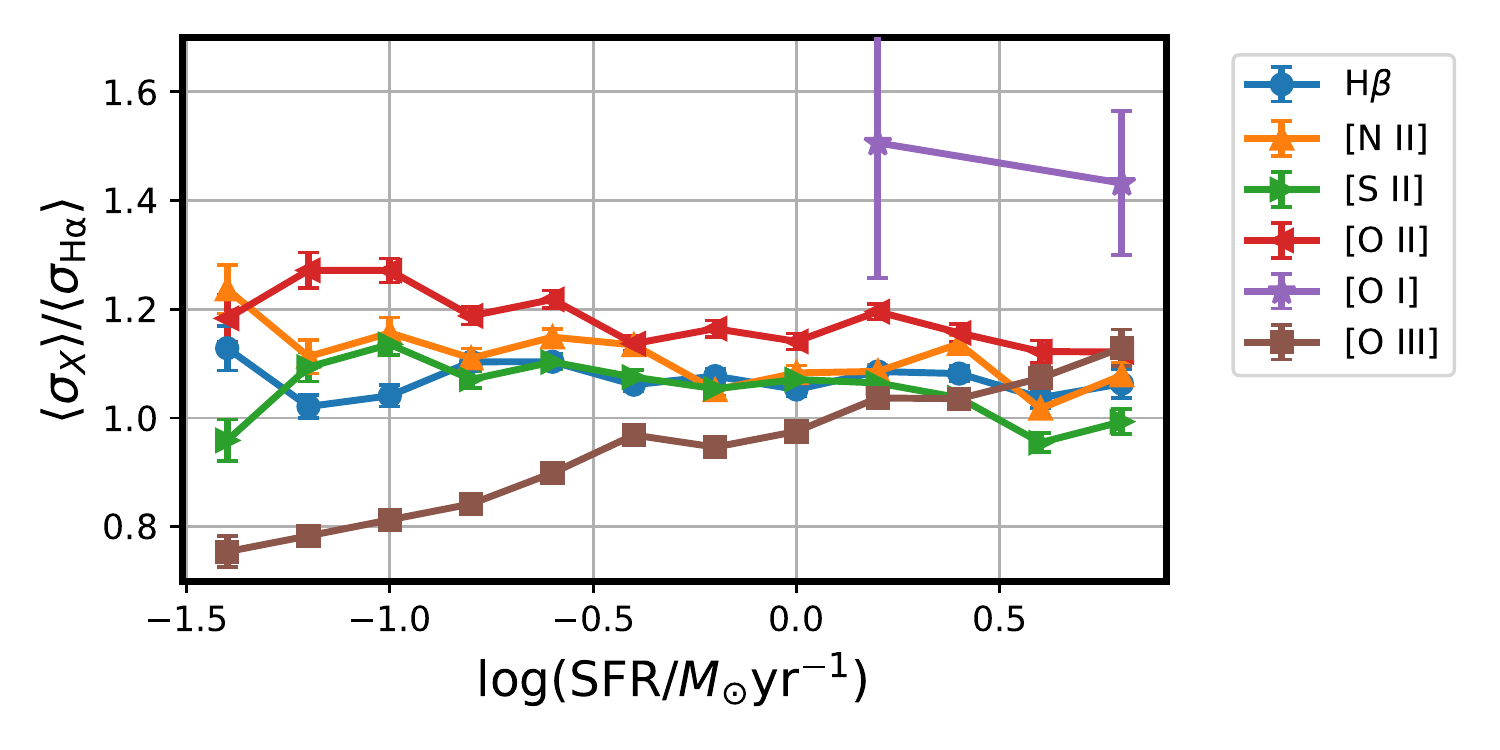}
\caption{Ratio between the intensity-weighted mean velocity dispersion of \Ha\ and
a variety of other strong emission lines as a function of the total \Ha-derived SFR.
Colored lines and points show the $2.5$-sigma clipped averages for the MaNGA galaxy
sample; error bars represent the uncertainty in the mean.
}
\label{paper3_multlines.fig}
\end{figure*}

One way in which we can assess the physical validity of this assumption is to
look at the corresponding velocity dispersion of other elements in the MaNGA data
as each will have unique properties.
In Figure \ref{paper3_multlines.fig} we plot the observed velocity dispersion
of the \Hb, \stwo\ $\lambda 6718 + 6732$, \ntwo\ $\lambda 6585$,
\oone\ $\lambda 6302$, \otwo\ $\lambda 3727 + 3729$,
and \othree\ $\lambda 5007$ emission lines relative to that of \Ha.\footnote{In deriving these values we have followed a similar procedure as in our calculation
of the \Ha\ velocity dispersion, restricting the sample to spaxels for which
the relevant emission line is detected at SNR $> 50$ and computing the intensity-weighted
mean velocity dispersion for a given galaxy according to the relevant line flux.}

We note that $\langle \sigma_{\Hb} \rangle$, $\langle \sigma_{\stwo} \rangle$, and $\langle \sigma_{\ntwo} \rangle$, and $\langle \sigma_{\otwo} \rangle$ all closely track $\langle \sigma_{\Ha} \rangle$, albeit offset
to larger values by about 5-10\%.\footnote{$\langle \sigma_{\otwo} \rangle$ is offset by  10-20\%, possibly because of observational bias from the rapidly-changing MaNGA LSF at wavelengths shortward of 4000 \AA.}
This offset was previous noted in \citet[][see their Figure 16]{paper1} at $\langle \sigma_{\Ha} \rangle$ values characteristic of
gas ionized by HII regions and may be due to the corresponding selection bias in
favor of only the brightest HII regions (since requiring \Hb\ SNR $> 50$ corresponds
to a selection cut \Ha\ SNR $\geq 150$).
$\langle \sigma_{\othree} \rangle$ in contrast shows a markedly different behavior,
rising steadily from $\sim 75$\% of $\langle \sigma_{\Ha} \rangle$ at the lowest
SFR to roughly equal values at the highest SFR.

This difference is unlikely to be due to any variation in the distribution of HII
regions within the galaxy giving rise to \Ha\ vs \othree\ emission.  Similarly, it
should not be an artifact of dust obscuration (since the trend is not observed in 
\Hb), or of differences in the atomic mass-dependent thermal broadening (since the
O$^{++}$ and N$^+$ ions have comparable mass, and S$^+$ is even more massive).
Rather, this is likely due to the significantly larger 35.2 eV ionization potential
of the O$^{++}$ ion compared to the $\sim 14$ eV ionization potential of the other ions
in Figure \ref{paper3_multlines.fig}.  As a result of the higher ionization potential,
\othree\ -emitting gas is located at denser regions deeper within the HII nebula
\citep[see, e.g.,][]{byler17} for which the turbulent velocity dispersion $\sigma_{\rm Turb}$ is smaller as the ionization front in HII regions generally
increases in speed with radius as the density drops \citep[see review by][]{osterbrock06}.

Indeed, our finding that $\langle \sigma_{\othree} \rangle < \langle \sigma_{\Ha} \rangle$ for the MaNGA data is in keeping with well established trends in
giant extragalactic HII regions, for which studies from \citet{hippelein86}
to \citet{bresolin20} have noted that $\langle \sigma_{\othree} \rangle$ is
systematically $\sim 2$ \kms\ less than $\langle \sigma_{\Ha} \rangle$.
The change in $\langle \sigma_{\othree} \rangle / \langle \sigma_{\Ha} \rangle$
that we observe over the range of total SFR probed by the MaNGA sample
of galaxies on the star-forming main sequence
may be a product of the variable size of the $O^{++}$ gas as a function
of both age and metallicity of the HII region.  However, it is perhaps
more likely that this trend is an artifact
of the increasing velocity dispersion $\sigma_{\rm Mol}$ between individual
HII regions in galactic disks.  As we show in \S \ref{molec.sec},
at low SFR the observed \Ha\ velocity dispersion $\sigma_{\Ha}$ is dominated
by the internal HII region dispersions $\sigma_{\rm Therm}$ and $\sigma_{\rm Turb}$,
and differences in $\sigma_{\rm Turb}$ due to stratification of different ionization
species within HII regions is therefore noticeable in the measured MaNGA velocity
dispersions.  In contrast, at higher SFR $\sigma_{\Ha}$ is dominated by
$\sigma_{\rm Mol}$, and small differences due to varying $\sigma_{\rm Turb}$ are
impossible to discern.

Our observation of \oone\ $\lambda 6302$ may be in keeping with this theory as well, as 
we find that $\langle \sigma_{\oone} \rangle / \langle \sigma_{\Ha} \rangle \approx 1.5$, consistent
with \oone\ production by neutral gas far out in the nebula.
However, \oone\ is significantly fainter than \Ha, and our requirement
of SNR $> 50$ thus limits our sample to just 400 spaxels across 25 galaxies.
We are thus unable to conclusively assess any potential trends in 
$\langle \sigma_{\oone} \rangle / \langle \sigma_{\Ha} \rangle$ with galactic SFR.

\subsection{Properties of the Molecular Gas}
\label{molec.sec}

Assuming that the corrections for thermal velocity dispersion and turbulence within
individual HII regions from \S \ref{hiiphysics.sec} are (on average) isotropic, we
can estimate the vertical velocity dispersion of the disk molecular gas in which individual HII
regions are embedded (many of which will typically fall within a $\sim$ kpc-sized
MaNGA spatial resolution element) as
\begin{equation}
    \sigma_{\rm Mol}^2 = \sigma_{\rm z}^2 - \sigma_{\rm Therm}^2 - \sigma_{\rm Turb}^2
\end{equation}
where $\sigma_z$ is the total vertical velocity dispersion computed following 
Equation \ref{sigz.eqn} with the best-fit parameters for the velocity ellipsoid
derived in \S \ref{inclin.sec}.

\begin{figure*}[!]
\epsscale{1.2}
\plotone{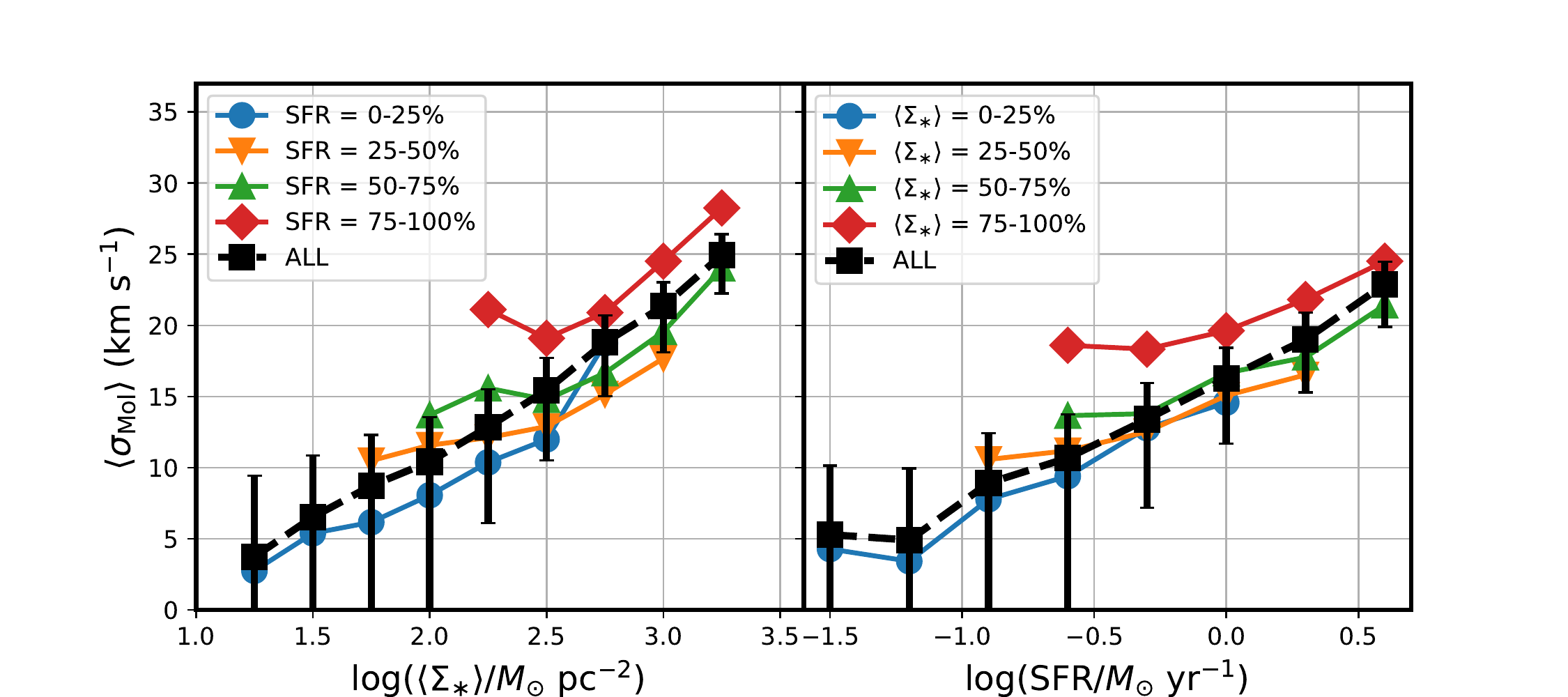}
\caption{Velocity dispersion $\sigma_{\rm Mol}$ of the molecular gas
(computed as the quadrature difference between the observed ionized gas velocity dispersion
and the estimated internal velocity dispersion of the HII regions) as a function
of the total SFR and the average disk stellar mass surface density.  
Colored lines and symbols show the
trend for four quartiles in total SFR, while the black lines and symbols show
the trend for all MaNGA galaxies.  Each point represents the $2.5\sigma$-clipped
average of the individual MaNGA data points.  Error bars on black point show estimated uncertainty in the mean assuming
that the turbulent velocity dispersion is $\sigma_{\rm Turb} = 10 \pm 5$ \kms.
All points represent $2.5\sigma$-clipped means applied to the observational sample.
}
\label{paper3_disksd.fig}
\end{figure*}

In Figure \ref{paper3_disksd.fig} we plot $\sigma_{\rm Mol}$ for the MaNGA galaxy sample
as a function of
the total SFR and the average stellar mass surface density (i.e., one potential
candidate for a global disk property that might be expected to scale with the vertical disk velocity dispersion).\footnote{Stellar
mass surface density $\Sigma_{\ast}$ is computed as the total stellar mass from the
NSA catalog, divided by the face-on disk surface area implied by the NSA
elliptical Petrosian radius fit to the observed broadband SDSS imaging data.}
These values are extremely uncertain; as illustrated by Figure \ref{masssfr.fig}
(lower-right panel) the observed velocity dispersion in the lowest SFR bin
is approximately 15 \kms, which is statistically indistinguishable from our assumed
14 \kms\ (see \S \ref{hiiphysics.sec}) due to thermal and turbulent motions within individual HII regions.  We cannot therefore say with confidence whether
$\sigma_{\rm Mol} = 2$ \kms\ in the lowest SFR (or $\langle \Sigma_{\ast} \rangle$) bin
as indicated by Figure \ref{paper3_disksd.fig} or some other small single-digit
value.  Regardless, under our present assumptions it appears to be the case that
in main-sequence galaxies with the lowest SFR
the observed ionized gas velocity dispersion is consistent with
being entirely produced by physics internal to individual HII regions, while at
the highest SFR the observed dispersion is dominated by the velocity dispersion
of the molecular gas in which those HII regions are embedded.

Despite the large uncertainties, our estimated $\sigma_{\rm Mol}$ are broadly 
consistent with estimates of the HI and molecular gas velocity dispersions available in the
literature.
\citet{ian12} for instance reports H I velocity dispersions of 7-17 \kms\ for 34
galaxies in the THINGS survey (which they ascribe to two components due to the
cold and warm neutral medium respectively).  
These values are close to
the 5-20 \kms\ range that we see in Figure \ref{paper3_disksd.fig}
for the log(SFR$/M_{\odot}$ yr$^{-1}$) $= -1.5$ to 0.5 range of the THINGS survey
\citep{walter08}.
Likewise, \citet{wilson19} observed that the CO-based velocity dispersion in local
(U)LIRGs increased as the square root of the molecular gas surface density.  Although
we cannot estimate a meaningful slope to this relation ourselves due to the uncertainty
of our correction for thermal and turbulent pressure within the HII regions,
we note that we are nonetheless consistent
with such a scaling relation- over a factor of 100 in $\Sigma_{\ast}$ Figure
\ref{paper3_disksd.fig} suggests that $\sigma_{\rm Mol}$ changes by about
a factor of ten.

\begin{figure}[!]
\epsscale{1.2}
\plotone{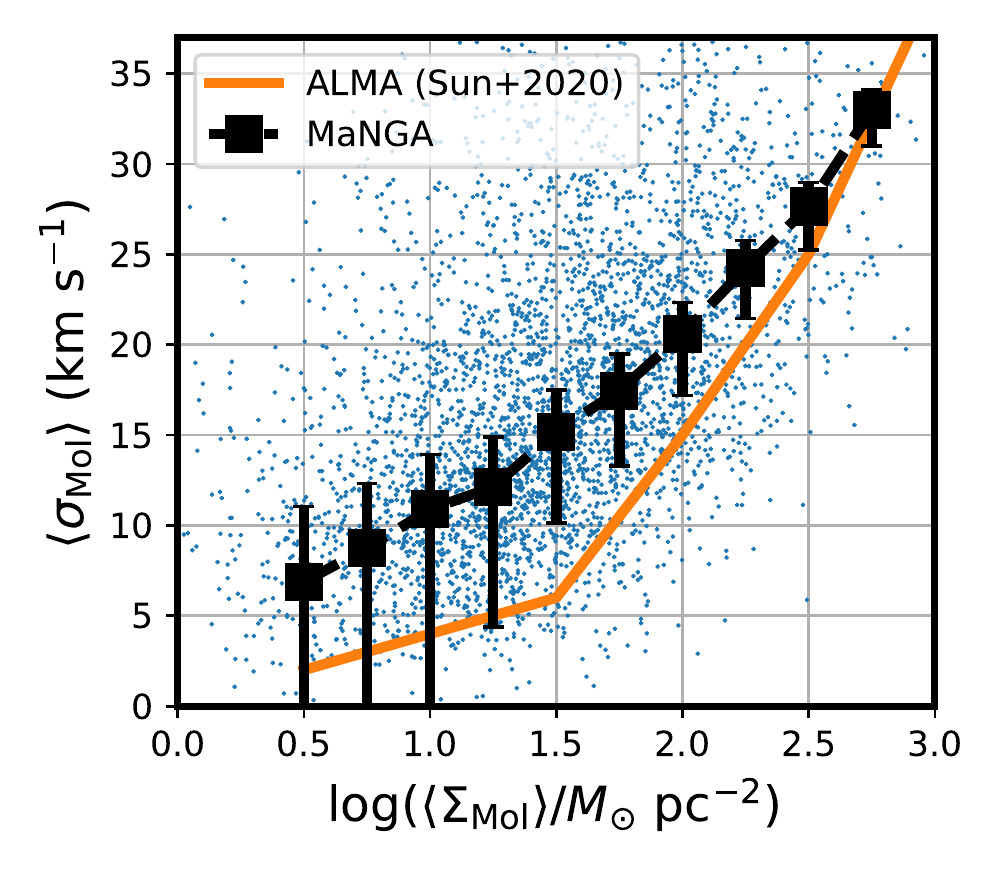}
\caption{Estimated molecular gas velocity dispersion as a function of the
mean molecular gas surface density for the MaNGA data (small blue points), along with 2.5-$\sigma$ clipped
mean values (solid black points).  Error bars represent the uncertainty in the mean, which is dominated by the
uncertainty in our adopted correction for HII region turbulence $\sigma_{\rm Turb}$.  The solid orange line represents the mean observed for 150-pc scale CO PHANGS-ALMA data by 
\citet{sun20}.
}
\label{paper3_molecular.fig}
\end{figure}

Molecular gas velocity dispersions have been studied in detail in recent years as well.
\citet{hughes13} for instance measured the 53-pc matched resolution properties of thousands
of molecular clouds in M51, M33 and the LMC and found a range of values from 3-15 \kms.
More recently, \citet{sun20} studied a sample of 70 nearby galaxies with the PHANGS-ALMA
CO survey composed of $> 100,000$ independent sightlines.
As these authors demonstrated (see their Fig 1), $\sigma_{\rm Mol}$ on 150pc scales
is a strong function of the molecular gas surface density, ranging from  $\sigma_{\rm Mol} = 1$ \kms\
to values in excess of 30 \kms\ at the highest surface densities.
We endeavor to compare our results quantitatively by estimating the average molecular gas
surface density of our MaNGA galaxies.  We do this by scaling our galaxy-averaged SFR surface densities
to $\Sigma_{\rm Mol}$ using the observed mean relation
for nearby disk galaxies from the HERACLES CO survey \citep[][their \Ha$+24\micron$ relation.]{leroy13}.
In Figure \ref{paper3_molecular.fig} we plot $\sigma_{\rm Mol}$ as a function of $\Sigma_{\rm Mol}$
for the MaNGA data in comparison to the median of the \citet[][see their Fig. 1]{sun20} data; while
the overall normalization of both axes is somewhat uncertain, the two data sets more or less agree that
$\sigma_{\rm Mol}$ increases rapidly over the range in molecular gas surface densities probed by
galaxies on the main sequence.
\citet{sun20} in particular noted that their results (and by extension ours as well) implied that the molecular
gas exceeds its self-gravitational binding energy by a small factor.

On its own, the MaNGA data is unable to conclusively determine whether or not there is a floor to the cold
gas velocity dispersion around 10 \kms\ as predicted by some galactic disk models \citep[e.g., the Transport+Feedback model of][]{krumholz18}.
As illustrated by Figure \ref{paper3_disksd.fig}, the presence or absence of such a floor is predicated on the
nature of the corrections for $\sigma_{\rm Therm}$ and $\sigma_{\rm Turb}$; if $\sigma_{\rm Turb} \geq 10$ \kms\ there
is no such floor and the observed \Ha\ velocity dispersions appear to be explained entirely by HII region physics.
If $\sigma_{\rm Turb} = 5$ \kms\ though, the error bars suggest that the MaNGA data is compatible with a 10 \kms\ floor.
Given the \citet{sun20} ALMA results though, we favor the former explanation in which $\sigma_{\rm Mol}$ continues
to decline towards lower SFR.

\begin{figure}[!]
\epsscale{1.2}
\plotone{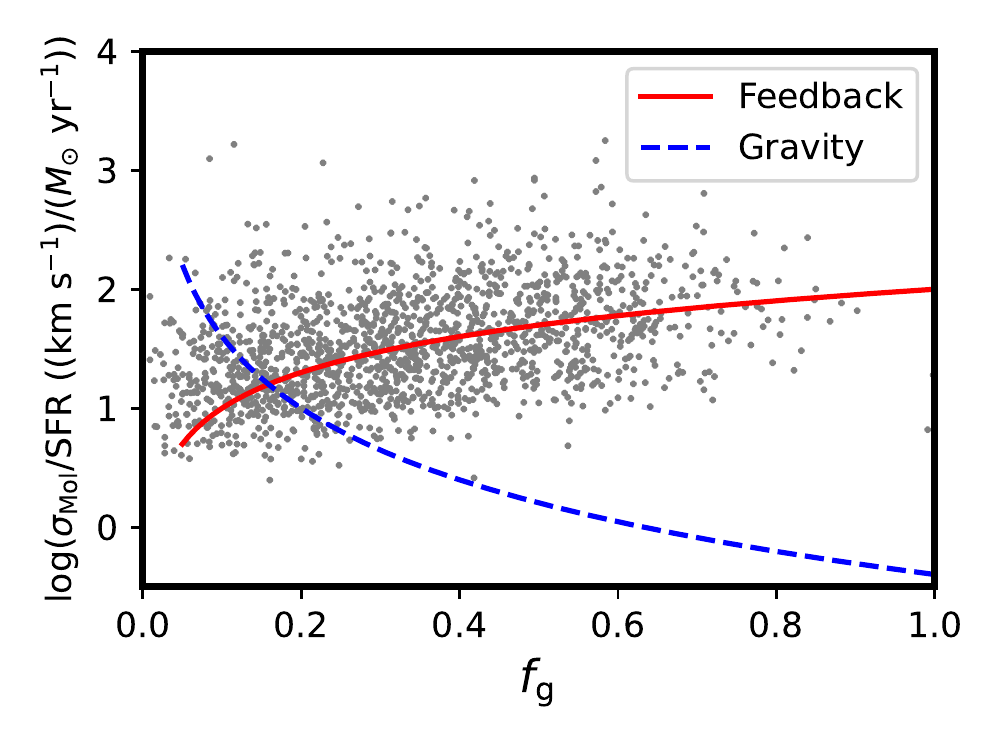}
\caption{Estimated mean molecular gas velocity dispersion divided by the total SFR as a function of the atomic gas mass fraction estimated
from HI observations for the MaNGA galaxy sample (grey points).  Overplotted for comparison are the feedback-based
and gravity-based turbulence models (SFR $\propto \sigma_{\rm Mol}^2$ and SFR $\propto f_g^2 \sigma_{\rm Mol}$ respectively) described by \citet{krumholz16}; note that the normalization of the models has been
chosen arbitrarily.
}
\label{krum16.fig}
\end{figure}

Theoretical models of the turbulence tend to fall into two broad groups; those in
which the turbulence is driven by stellar feedback \citep[e.g.,][]{ostriker11,fg13}, and those in which the turbulence is driven by
gravitational instabilities \citep[e.g.,][]{krumholz10}.  Based on the work of \citet{krumholz16}, in the first scenario 
we would expect that the SFR $\propto \sigma_{\rm Mol}^2$, while the latter would predict that SFR $\propto f_g^2 \sigma_{\rm Mol}$ where $f_g$ is the gas fraction.  We evaluate these two possibilities in Figure \ref{krum16.fig} by plotting
$\sigma_{\rm Mol}$ divided by the SFR for the MaNGA galaxies as a function of the gas fraction.
In the feedback driven model, $\sigma_{\rm Mol}/SFR \propto SFR^{-1/2}$ with no explicit dependence on the gas fraction.
Based on the HI-MaNGA data though we note that in our galaxies SFR $\propto f_g^{-2}$ (i.e., total SFR is highest
in the highest-mass galaxies for which the average atomic gas fraction is low), implying that $\sigma_{\rm Mol}/SFR \propto f_g$.
Such a relation is exactly what we see in our data.  In contrast, the gravity driven model predicts
$\sigma_{\rm Mol}/SFR \propto f_g^{-2}$ which is strongly disfavored by the MaNGA observations.
At least in the range of conditions present in the $z=0$ galactic main sequence,
turbulence therefore appears to be consistent with feedback driven by the observed star formation rate.  We note that a similar conclusion was reached by \citet{bacchini20},
who found that the atomic and molecular gas turbulence in a sample of
ten nearby star forming galaxies was energetically consistent with supernova feedback
alone after accounting for increased dissipation timescales due to radial flaring of
the galactic gas disk.

Assuming that the molecular gas is in pressure equilibrium with the
vertical disk velocity dispersion balancing the gravitational force of the average
disk mass surface density it is possible to estimate
the effective scaleheight $h$ across which our HII regions are distributed.  Following traditional
derivations \citep[see, e.g.,][]{wilson19,jorge21}, we define
\begin{equation}
    h = \sigma_{\rm Mol}^2/2 \pi G \Sigma_{\rm Tot}
\label{scaleheight.eqn}
\end{equation}
where $\Sigma_{\rm Tot} = \Sigma_{\ast} + \Sigma_{\rm Mol}$.

For the range of values shown in Figures \ref{paper3_disksd.fig} and \ref{paper3_molecular.fig},
Equation \ref{scaleheight.eqn} implies 
disk scaleheights in the range $\sim 10-40$ pc.  While estimates
of the molecular gas scaleheight in the Milky Way and other galaxies vary substantially according to the tracer used (e.g., CO emission, \ctwo\ emission,
870 $\micron$ continuum emission, etc), our values are consistent with those
similarly sensitive to the actively star-forming layer;
\citet[][and references therein]{anderson19} for instance find a vertical scale height 
of $\sim 30$pc in the Milky Way based on the observed distribution of Galactic HII regions in the WISE catalog (ranging from 25 pc for the youngest to 40 pc for the oldest such regions).
Likewise, at a SFR of $1 M_{\odot}$ yr$^{-1}$, Figure \ref{paper3_disksd.fig} implies
a molecular gas velocity dispersion of $\sigma_{\rm Mol} = 15$ \kms, entirely in keeping
with the youngest, most metal-rich Galactic stellar populations which have vertical velocity dispersions in the range 10-20 \kms\ \citep{bovy12a, hayden17}.


\subsection{Diffuse Ionized Gas}

Thus far, we have concentrated our attention exclusively on the gas ionized by star-forming
processes in HII regions, and the corresponding implications for the underlying
molecular gas layer.
However, star formation is not the only source of H ionizing photons;
diffuse ionized gas (DIG)
can represent a significant component of the total \Ha\ emission
from a given galaxy
\citep{oey07,zhang17} and offers an additional means by which to study the structure
of the ionized gas beyond the confines of traditional HII regions.

As discussed in \citet{paper2}, DIG can be reliably identified by selecting spaxels
with low \Ha\ equivalent width that strongly cluster in the traditional
LI(N)ER region of the BPT diagram.  Using the
\stwo/\Ha\ and \othree/\Hb\ criteria established in \citet{paper2}, 
we select all DIG-like spaxels in our sample of otherwise
star-forming galaxies.  By definition, very few such spaxels meet our ideal
SNR $>50$ threshold required to minimize biases from the large MaNGA instrumental
LSF.  We therefore relax this criterion to require SNR $> 10$, resulting in a
sample of 39,000 spaxels distributed amongst 2935 individual galaxies.  While this
spaxel sample is just 3\% the sample size of star-forming spaxels, it is
nonetheless sufficient to draw some general conclusions.

\begin{figure}[!]
\epsscale{1.2}
\plotone{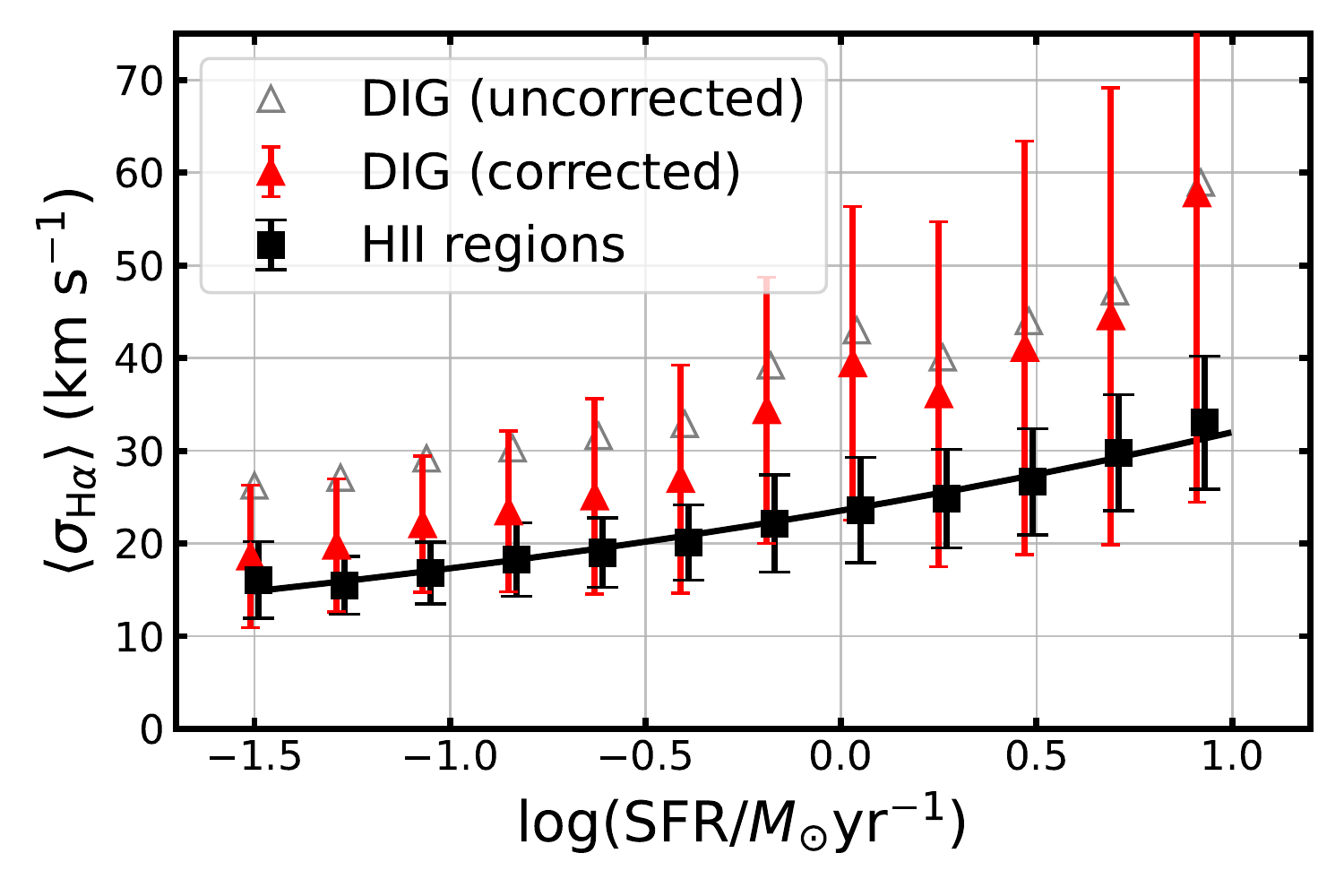}
\caption{Intensity-weighted mean velocity dispersion of the diffuse ionized
gas within the MaNGA star-forming galaxy sample as a function of the total star-formation rate (derived from the HII regions).  Open grey triangles show the
binned data for the $\sim 39,000$ spaxels across 2935 individual galaxies meeting
our selection criteria.  This relation steepens significantly after application
of a correction to account for low values lost due to the MaNGA instrumental
LSF (solid red triangles).  Also shown for comparison are the velocity dispersions
for HII regions (filled black squares).  Error bars represent the observed width of
the galaxy distribution corrected for artificial width due to uncertainties in the
instrumental LSF.
All points represent $2.5\sigma$-clipped means applied to the observational sample.
}
\label{paper3_dig.fig}
\end{figure}

In Figure \ref{paper3_dig.fig} (open triangles)
we plot the intensity-weighted mean \Ha\ velocity dispersion of these 2935 MaNGA
galaxies as a function of the total galaxy SFR (as estimated from the star-forming
spaxels).  Unlike for the star-forming spaxels, the sample is strongly biased
towards the lowest \Ha\ SNR in the sample, and we must therefore correct the intensity-weighted mean values for survival statistics
from the large MaNGA instrumental LSF
following Figure 15 of \citet{paper1}.
After such a correction (filled red triangles in Figure \ref{paper3_dig.fig}) we note
that the velocity dispersion of the DIG in the lowest-SFR systems is nearly
indistinguishable from the velocity dispersion of the HII regions, but at the highest
SFR is nearly double that of the HII regions.

We remarked in \citet{paper2} that the velocity dispersion of the DIG was larger in general
than the velocity dispersion of HII regions; Figure
\ref{paper3_dig.fig} breaks this down into a physical picture of the evolving
sources of the DIG with increasing galactic SFR.  At low SFR (and low masses)
in which there is not a significant evolved stellar population the DIG
may be predominantly created by leakage from HII regions, and the overall velocity
structure of the HII and DIG gas is thus similar.  At higher SFR (and high masses)
the galactic stellar disk is much more massive and well-established, and 
we may be observing the dominant source of the DIG-illuminating photons
shifting to hot evolved stars distributed in a much thicker (and higher velocity dispersion) disk.

Such a result is broadly compatible with the observational results of 
\citet{db20} who noted that the velocity dispersion of the DIG was measurably
larger than that of the HII regions at 10pc scales in NGC 7793 observed with the MUSE
integral field spectrograph, and of \citet{denbrok20} who reached similar conclusions
based on relative asymmetric drift measurements for 41 galaxies observed
with MUSE.


\section{Summary}
\label{summary.sec}

We have used the completed MaNGA survey to study the behavior of \Ha\ velocity
dispersions for 4517 star-forming
galaxies at $z \sim 0.02$ that sample the star-forming main sequence
from $M_{\ast} = 10^9 - 10^{11} M_{\odot}$.
Despite the large instrumental line spread function, our detailed understanding of systematics
allows us to reliably measure ensemble velocity dispersions down to 15-30 \kms\
characteristic of the galactic ionized gas disk.
We summarize our main conclusions as follows:

\begin{itemize}
    \item There are strong, well-defined correlations between both the localized SFR surface density $\Sigma_{\rm SFR}$ and the local ionized gas velocity dispersion $\sigma_{\Ha}$, and between the total galactic SFR and the intensity-weighted mean velocity dispersion $\langle \sigma_{\Ha} \rangle$.  In the latter case, $\langle \sigma_{\Ha} \rangle$ increases from $16.1 \pm 4.1$ \kms\ at SFR of 0.03 $M_{\odot}$ yr$^{-1}$ to
    $33.0 \pm 7.2$ \kms\ at SFR of $8 M_{\odot}$ yr$^{-1}$.
    Our results in both cases are consistent with previous measurements from smaller samples of galaxies at higher spectral resolution.

    \item Using the statistical power of MaNGA to control for a variety of sub-populations, we have demonstrated that trends between ionized gas velocity dispersion $\Sigma_{\rm SFR}$ and stellar 
    mass $M_{\ast}$ are subdominant to the relation with total SFR.  That is, total SFR is the strongest driver of the trend in velocity dispersions,
    and apparent trends with $\Sigma_{\rm SFR}$ and $M_{\ast}$ are produced by the correlation of these quantities with total SFR.

    \item We have used velocity dispersions derived from multiple nebular emission lines \Ha, \Hb, \ntwo, \stwo, and \othree\ to constrain our understanding of the ionized gas physics internal to the HII regions that are responsible for the majority of the observed emission.  We find that
    the \othree\ velocity dispersion is systematically smaller than the \Ha\ velocity dispersion, consistent with models of HII regions in which
    the turbulent velocity dispersion $\sigma_{\rm Turb}$ increases with radius within the HII region as the density and ionization parameter decrease.
    
    \item Assuming a model for the thermal and turbulent contributions of individual HII regions ($\sigma_{\rm Therm}$ and $\sigma_{\rm Turb}$ respectively), we have estimated the velocity dispersion $\sigma_{\rm Mol}$ of the molecular gas within which the HII regions in our sample
    are embedded.  We find that $\sigma_{\rm Mol}$ increases from $\sim 5$ \kms\ at low SFR to $\sim 30$ \kms\ at high SFR.  Casting our result
    in terms of the galaxy-averaged cold gas fraction, these results agree closely with recent molecular gas observations from the PHANGS survey.
    
    \item The velocity dispersion of the diffuse ionized gas (i.e., \Ha\ emission produced by regions with nebular line ratios inconsistent
    with star-formation models) is comparable to that of the HII-region ionized gas at low SFR and increases more rapidly to $\sim 60$ \kms\ in the high-SFR subsample.  This may be consistent with a transition from DIG produced mostly by HII-region leakage at low SFR to ionization from hot evolved stars in the more massive stellar disks present at high SFR.
    
    \item Using positional information of the spaxels that compose each galaxy, we have assessed the mean velocity dispersion as a function of galaxy inclination $i$ and azimuthal angle $\phi$ within the galaxy.  The observed relation suggests that the ionized gas has a velocity dispersion ellipsoid in which the radial and azimuthal velocity dispersions are appreciably larger than the vertical velocity dispersion ($\sigma_z/\sigma_r = 0.84 \pm 0.03$ and
    $\sigma_{\phi}/\sigma_r = 0.91 \pm 0.03$).  These ratios differ from unity in the same sense as for the youngest stellar population in the Milky Way, suggesting that young stars and the birth clouds that they come from have similar large-scale influences on their orbital dynamics
    (e.g., bars and spiral arms within disks).
\end{itemize}

Overall, the MaNGA data are consistent with a picture in which HII regions
condense within localized overdensities in galactic molecular disks.
As these HII regions evolve the stars within them, newly freed from the
molecular gas disk, gradually diffuse to larger
scaleheights as they relax into the gravitational potential of the stellar disk.
In the Milky Way, the most metal-rich such stellar populations have
scaleheights of order 200 pc,
increasing to 400-600 pc for progressively older and more metal-poor populations \citep{bovy12b}.
At higher redshifts $z \sim 2$, the ubiquitously large ionized gas velocity dispersions
$\sim 70$ \kms\ \citep[e.g.][]{law09,ubler19} may imply a similarly large dispersion
between HII regions embedded in the molecular gas, fueling the correspondingly
thick stellar disk scaleheights that are observed \citep{law12morph,zhang19}.
Since main-sequence star-forming galaxies at such redshifts tend to have SFR
$\sim 10-100 M_{\odot}$ yr${-1}$ \citep[e.g.,][]{wuyts11,theios19} typically only seen in
(U)LIRGs and similar systems in the nearly universe, the overall picture may be consistent with
`upside-down' disk formation in which thick disks form early and thin disks
form at later times \citep[e.g.,][]{yu21}.

\begin{acknowledgments}

We thank D. Andersen for access to the \ntwo\ DensePak measurements used in this work. MAB acknowledges support from NSF/1814682.  DRL appreciates precise and insightful comments from the anonymous referee which improved the final version of this manuscript.

Funding for the Sloan Digital Sky Survey IV has been provided by the Alfred P. Sloan Foundation, the U.S. Department of Energy Office of Science, and the Participating Institutions. SDSS-IV acknowledges
support and resources from the Center for High-Performance Computing at
the University of Utah. The SDSS web site is www.sdss.org.

SDSS-IV is managed by the Astrophysical Research Consortium for the 
Participating Institutions of the SDSS Collaboration including the 
Brazilian Participation Group, the Carnegie Institution for Science, 
Carnegie Mellon University, the Chilean Participation Group, the French Participation Group, Harvard-Smithsonian Center for Astrophysics, 
Instituto de Astrof\'isica de Canarias, The Johns Hopkins University, Kavli Institute for the Physics and Mathematics of the Universe (IPMU) / 
University of Tokyo, the Korean Participation Group, Lawrence Berkeley National Laboratory, 
Leibniz Institut f\"ur Astrophysik Potsdam (AIP),  
Max-Planck-Institut f\"ur Astronomie (MPIA Heidelberg), 
Max-Planck-Institut f\"ur Astrophysik (MPA Garching), 
Max-Planck-Institut f\"ur Extraterrestrische Physik (MPE), 
National Astronomical Observatories of China, New Mexico State University, 
New York University, University of Notre Dame, 
Observat\'ario Nacional / MCTI, The Ohio State University, 
Pennsylvania State University, Shanghai Astronomical Observatory, 
United Kingdom Participation Group,
Universidad Nacional Aut\'onoma de M\'exico, University of Arizona, 
University of Colorado Boulder, University of Oxford, University of Portsmouth, 
University of Utah, University of Virginia, University of Washington, University of Wisconsin, 
Vanderbilt University, and Yale University.
\end{acknowledgments}

\end{document}